\documentclass{elsart}

\journal{astro-ph: Review Article}

\usepackage{epsfig}

\newcommand{\be}{\begin{eqnarray}}
\newcommand{\ee}{\end{eqnarray}}
\def\lsim{\mathrel{\rlap{\lower3pt\hbox{\hskip1pt$\sim$}}
    \raise1pt\hbox{$<$}}} 
\def\gsim{\mathrel{\rlap{\lower3pt\hbox{\hskip1pt$\sim$}}
    \raise1pt\hbox{$>$}}} 
\newcommand{\msun}{M_\odot}
\newcommand{\rsun}{R_\odot}
\newcommand{\BH}{\rm BH}
\newcommand{\He}{\rm He}

\def\bi{\begin{enumerate}}
\def\ei{\end{enumerate}}

\def\msun{M_\odot}

\def\kpc{kpc}

\begin{document}

\runauthor{Lee \& Brown}

\begin{frontmatter}
\title{ON THE THEORY OF GAMMA RAY BURSTS AND HYPERNOVAE:\\
The Black Hole Soft X-ray Transient Sources}

\author[snu,suny]{Chang-Hwan LEE\thanksref{chl}}
\author[suny]{Gerald E. BROWN\thanksref{geb}}

\address[snu]{BK21 Physics Research Division and Center for Theoretical 
Physics,\\ School of Physics,
Seoul National University, Seoul 151-747, Korea}
\address[suny]{Department of Physics \& Astronomy,
        State University of New York\\
        Stony Brook, New York 11794, USA}
\thanks[chl]{chlee@phya.snu.ac.kr}
\thanks[geb]{popenoe@nuclear.physics.sunysb.edu}

\begin{abstract}
We show that a common evolutionary history can produce
the black hole binaries in the Galaxy in which the
black holes have masses of $\sim 5-10\msun$. 
In with low-mass, $\lsim 2.5\msun$, ZAMS (zero age main sequence) companions,
the latter remain in main sequence during the active stage
of soft X-ray transients (SXTs), most of them being of K or M
classification. In two intermediate cases, IL Lupi and
Nova Scorpii with ZAMS $\sim 2.5\msun$ companions
the orbits are greatly widened because of large mass loss
in the explosion forming the black hole, and whereas these
companions are in late main sequence evolution, they are
close to evolving. Binaries with companion ZAMS masses
$\gsim 3\msun$ are initially ``silent" until the companion
begins evolving across the Herzsprung gap.

We provide evidence that the narrower, shorter period binaries,
with companions now in main sequence, are fossil
remnants of gamma ray bursters (GRBs). We also show that the
GRB is generally accompanied by a hypernova explosion
(a very energetic supernova explosion).
We further show that the binaries with evolved companions are
good models for some of the ultraluminous X-ray sources (ULXs)
recently seen by Chandra in other galaxies.

The great regularity in our evolutionary history, especially
the fact that most of the companions of ZAMS mass $\lsim 2.5\msun$
remain in main sequences as K or M stars can be explained by
the mass loss in common envelope evolution to be Case C;
i.g., to occur only after core He burning has finished.
Since our argument for Case C mass transfer is not generally 
understood in the community, we add an appendix, showing that with
certain assumptions which we outline we can reproduce the
regularities in the evolution of black hole binaries by Case C
mass transfer.
\end{abstract}


\end{frontmatter}

\def\BH{{\rm BH}}
\def\SN{{\rm SN}}
\def\L{{\rm L}}
\def\I{{\rm I}}
\def\II{{\rm II}}
\def\D{{\rm D}}
\def\Sch{{\rm Sch}}


\tableofcontents


\section{INTRODUCTION}
\label{intro}

The discovery of afterglows to gamma-ray bursts (GRBs) 
has greatly increased the
possibility of studying their physics. 
Recent observations strongly suggest a connection between GRBs
and supernovae, with indication that the supernovae in 
question are especially energetic and of type Ib/c, i.e., core
collapses of massive stars which have lost their hydrogen envelope
(see Van Paradijs et al.\cite{paradijs00}, and references therein).
This supports suggestions by 
Woosley\cite{Woosley93B} and Paczy\'nski\cite{paczynski98}
for the origin of GRBs in stellar core collapses. 
The hydrodynamics of a jet escaping from a star and causing its explosion
was explored in detail by MacFadyen \& Woosley\cite{MacFadyen99}, 
who showed that
contrary to accepted wisdom, a fairly baryon-free, ultra-relativistic 
jet could plow through the collapsing star and emerge with large Lorentz
factors. The powering of the outflow by coupling of high magnetic fields
to the rotation of the black hole~\cite{bz77}, first
suggested by 
Paczy\'nski\cite{paczynski98} in the context of GRBs, was
worked out in detail by Van Putten\cite{putten99,putten01}.
Li has also discussed the deposition of energy from a black hole
into the accretion disk in a recent series of papers\cite{Li2000}.

Building on these thoughts, we have modeled both the powering of a
GRB by black-hole rotation and the stellar evolution pathways
that set up favorable conditions for that mechanism~\cite{grb2000}.
An essential ingredient in this model is a rapidly rotating black hole,
and it is this aspect that we focus on in the present paper.

A massive star in
a close binary will spin faster than a massive single star
for a number of reasons: first, when the
hydrogen envelope is lifted off by spiral-in, it will cease to serve as a 
sink of angular momentum for the core. Second, the tidal friction
concomitant with the spiral-in process will spin up the inner region,
giving it a larger angular momentum than the same region in a 
single star~\cite{Rasio96}.
Third, tidal coupling in the close binary will tend to bring the primary
into corotation with the orbital period. 
This latter process is not very efficient in
the short post spiral-in life of the binaries we consider, but its effect
does probably matter to the outer layers of the helium star, which can be
important for our work. With its more rapid rotation, the helium star
then forms a black hole with a large Kerr parameter, which immediately
after its formation (in a few seconds) begins to input power into its
surroundings at a very high rate. This, then, powers both a 
GRB~\cite{grb2000} and the expulsion of the material that was
centrifugally prevented from falling into the black hole. In fact,
Van Putten\cite{putten99,putten01} estimates that the power input 
into that material
exceeds that into the GRB and Li\cite{Li2000b} also finds that more energy
can be extracted by the disk than by the GRB. 
It should be noted that an initially less
rapidly rotating black hole could be spun up by disk accretion quite
rapidly, and start a similar process after some accretion has taken
place~\cite{MacFadyen99,grb2000}. Some implications
of such more complicated sequences of events are discussed by 
Lee et al.\cite{Lee2001}.

More than dozen soft X-ray transient (SXT) black hole binaries 
were observed in our Galaxy.
By far the most famous one is Nova Scorpii 94. 
Israelian et al.\cite{Israelian99} found the overabundances of heavy 
elements in the subgiant companion star in Nova Scorpii, and suggested that
it is a relic of hypernova explosion. Observed high system velocity and
the quasi periodic oscillations support the hypernova
explosion associated with rapidly rotating black hole.
In a more recent analysis~\cite{Lee2002}, we found the correlation
between the black hole masses and the reconstructed
pre-explosion orbital periods of SXTs. In binaries
with preexplosion orbital period less than 12 hours,
rapidly rotating black holes are formed in the core of helium
stars by the prompt collapse, and the outer part of the helium star 
is held from immediate collapse by the augular momentum support.
Based on these, we suggest that the SXTs
with preexplosion orbital period less than or similar to 12 hours
are the relics of GRBs and the hypernovae.

Since the afterglows have thus
far only been seen for long GRBs (duration $\gsim 2$\,s), we
shall concentrate on the mechanism for this subclass. The shorter bursts
(duration $\lsim 2$\,s) may have a different origin; specifically, it
has been suggested that they are the result of compact-object mergers
and therefore offer the intriguing possibility of associated outbursts
of gravity waves. (Traditionally, binary neutron stars have been 
considered in this category~\cite{Eichler89,Janka99}.
More recently, Bethe \& Brown~\cite{Bethe98} have shown 
that low-mass black-hole, 
neutron-star binaries, which have a ten times greater formation rate and 
are stronger gravity-wave emitters, may be the more promising source of
this kind.)


The plan of this paper is as follows.
In Sec.~\ref{grbhypernova}, the observational indications of
the GRB and supernova/hypernova association are
discussed.

In Sec.~\ref{energy}, we give the simple argument for
the energetics of GRBs and Hypernovae. Our simple argument
is based on the Blandford-Znajek mechanism, and we discuss
various channels of the power output from the black hole-accretion
disk system.  Woosley's  Collapsar model is
discussed as a progenitor of GRBs and Hypernovae. 

In Sec.~\ref{sxts}, we summarize the observational
indications of SXTs, especially Nova Scorpii
which is by far the most interesting SXT.
We also discuss our recent discoveries on the empirical
correlation between the orbital periods and black hole masses
of X-ray transient black hole systems.

The evolution of the SXTs is discussed
in Sec.~\ref{evol}. Based on the Case C mass transfer,
which is essential part of the evolution, we could pin down
the common envelope efficiency.

Using the evolution scenario in the previous section, 
in Sec.~\ref{reconst},
we reconstructed the pre-explosion orbital period and
the black hole masses at the time of formation.
We found a regularity in the reconstructed mass-period
relations.

In Sec.~\ref{rotate}, we discuss a schematic model
which explains the correlation between the black hole masses
and the preexplosion periods. In SXTs with
short pre-explosion orbital periods, less than or similar
to 12 hours, rapidily rotating black holes are formed.
Based on these results, we suggest that the progenitors of
black holes in SXTs are the sources
of GRBs and Hypernovae if their preexplosion spin periods
are less than or similar to 12 hours. In this model we
assumed that the tidal interaction synchronized the
orbital period and spin period of the black hole progenitor.

In Sec.~\ref{other}, we discuss other observational issues
of SXTs. We give estimates on the chemical
composition of a few X-ray transients. We discuss
the untraluminous X-ray sources in our Galaxy. We also
discuss the population synthesis of SXTs and GRBs.

Our final discussion and conclusion follows in Sec.~\ref{conclu}.

In the Appendix, we discuss the Case C mass transfer in detail,
which is essential part of the evolution of the black hole progenitors.
We also show when there will be deviation from Case C, indicating that
these will not affect our main results.

\section{GRB AND SUPERNOVA/HYPERNOVA ASSOCIATION: OBSERVATIONS}
\label{grbhypernova}

An important recent clue to the origin of GRBs is the
probable association of some of them with ultra-bright 
type Ibc supernovae~\cite{Galama98C,Bloom99C,Galama00A,Price02}.  
Recently, there appeared a beautiful
example of the GRB-hypernova association~\cite{bloom2002}.
In this paper the light curve showed a bump, reddened by the
many metal lines, rising on that from the GRB afterglow
after several days. This bump was fit well by red shifting 
the known hypernova light curve from SN1998bw, previously
interpreted as a hypernova explosion, to the relevant
redshift (0.36). 
 
The very large explosion
energy implied by fitting the light curve of SN\,1998bw, 
which was associated with
GRB\,980425\cite{Galama98C}, indicates that a black hole was formed
in this event~\cite{Iwamoto98}. This provides
two good pieces of astrophysical information: it implicates
black holes in the origin of GRBs, and it demonstrates that
a massive star can explode as a supernova even if its core
collapses into a black hole.
 
The GRB and accompanying hypernova have 
a common explanation in the model of Brown et al.\cite{grb2000} 
and Lee et al.\cite{Lee2002}.
Both are powered by the rotational energy of the black hole,
the hypernova through the closed field lines coupling
the black hole to the accretion disk.
 
In a review by Tsvi Piran\cite{Piran02}
it is pointed out that the actual GRB energy is narrowly
distributed around a ``mere" $\sim 10^{51}$ ergs.
Hypernova calculations which reproduce the spectra of 1998bw
seem to need $\sim 10^{52}$ ergs, ten times more
(see the results in Fig.~4 of Brown et al.\cite{grb2000}.)
We have shown that the total energy from the rotational energy
of the black hole is $\sim 10^{53}$ ergs~\cite{grb2000,Lee2002}.

\section{ENERGETICS OF GRB AND HYPERNOVA}
\label{energy}

\subsection{Simple circuitry}

We start from the viewpoint that the GRB is
powered by electromagnetic energy extraction from a spinning black hole,
the so-called Blandford-Znajek\cite{bz77} mechanism.  This
was worked out in detail by Lee et al.\cite{Lee99A,Lee99B}, who
built on work by Thorne et~al.\cite{Thorne86} and Li\cite{Li00}.  
They have shown that with the circuitry
in a 3$+$1 dimensional description using the Boyer-Lindquist metric,
one can have a simple pictorial model for the BZ mechanism.

\begin{figure}
\centerline{\epsfig{file=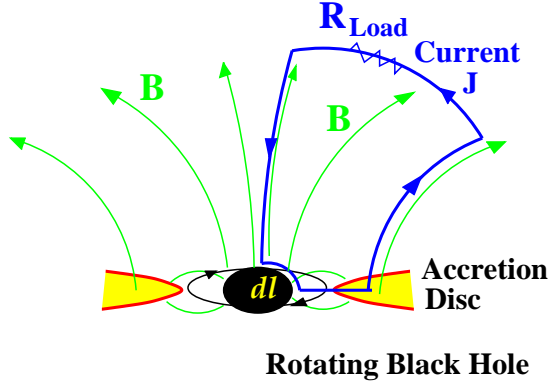,height=2.0in}} 
\caption{The
black hole in rotation about the accretion disk. A circuit,
in rigid rotation with the black hole is shown. This
circuit cuts the field lines from the disk as the black hole
rotates, and by Faraday's law, produces an electromotive force.
This force drives a current. More detailed discussion is given in
the text. 
} \label{bzcirc}
\end{figure}

Although our numbers are based on the detailed review of 
Lee et al.\cite{Lee99A},
which confirms the original Blandford-Znajek paper\cite{bz77},
we illustrate our arguments with the pictorial
treatment of Thorne et~al.\cite{Thorne86} in {\it ``The Membrane
Paradigm"}. Considering the time as universal in the
Boyer-Lindquist metric, essential electromagnetic and statistical
mechanics relations apply in their 3$+$1 dimensional manifold. We
summarize their picture in our Fig.~\ref{bzcirc}.

The simple circuitry which involves steady state current flow is,
however, inadequate for describing dissipation of the black hole
rotational energy into the accretion disk formed from the original
helium envelope. In this case the more rapidly rotating black hole
tries to spin up the inner accretion disk through the closed field
lines coupling the black hole and disk. Electric and magnetic fields vary
wildly with time. Using the work of Blandford \& Spruit\cite{Blandford00} 
we show that this dissipation occurs in an oscillatory fashion, giving
a fine structure to the GRB, and that the total dissipation should
furnish an energy comparable to that of the GRB to the accretion disk.
We use this energy to drive the hypernova explosion.

Not any black-hole system will be suitable for making GRB: the
black hole must spin rapidly enough and be embedded in a strong
magnetic field. Moreover, the formation rate must be high enough
to get the right rate of GRB even after accounting for substantial
collimation of GRB outflows. 
We argue
that the systems known as black-hole transients are the relics of
GRBs, and discuss the recent evidence from high space velocities
and chemical abundance anomalies that these objects are relics of
hypernovae and GRBs; we especially highlight the case of
Nova Scorpii 1994 (GRO\,J1655$-$40).

The surface of the black hole can be considered as a conductor with surface
resistance $R=4\pi/c=377$ ohms. A circuit that rotates rigidly
with the black hole can be drawn from the loading region, the
low-field region up the axis of rotation of the black hole in
which the power to run the GRB is delivered, down a magnetic field
line, then from the North pole of the black hole along the
(stretched) horizon to its equator. From the equator we continue
the circuit through part of the disk and then connect it upwards
with the loading region. We can also draw circuits starting from
the loading region which pass along only the black hole or go
through only the disk, but adding these would not change the
results of our schematic model.

Using Faraday's law,
    \be
    V =\int
    [\vec v\times\vec B]\cdot d{\vec l},
    \ee
we can integrate downwards along our current loop. Because this law
involves $\vec v\times \vec B$ the integrals along the field lines
make no contribution. We do get a contribution $V$
from the integral from North
pole to equator along the black hole surface. Further
contributions to $V$ will come from cutting the field lines from
the disk. We assume the field $B$ to be weak enough in the loading
region to be neglected.

The GRB will be powered by
   \be
   \dot E_{\rm GRB}=I_{\BH+\D}^2 R_\L
   \ee
where $R_\L$ is the resistance of the loading region, and
   \be
   I_{\BH+\D}^2 =\left(\frac{V_\D+V_{\BH}}{R_\D+R_{\BH}+R_\L}\right)^2.
   \ee

The load resistance has been estimated in various ways and for various
assumptions~\cite{Lovelace79,MacDonald82,Phinney83}.
All estimates agree that to within a factor
of order unity $R_\L$ is equal to $R_\BH$.\footnote{
  Whether this is really so, or just wishful thinking is not clear to us.
  It is well known that such impedance matching maximizes the power
  output.}
The disk resistance $R_D$ is usually omitted.

In a similar fashion, some power will be deposited into the disk
    \be
    \dot E_{disk}=I_{BH+D}^2 R_D
    \ee
but this equilibrium contribution will be small because of the low
disk resistance $R_D$.

\begin{figure}
\centerline{\epsfig{file=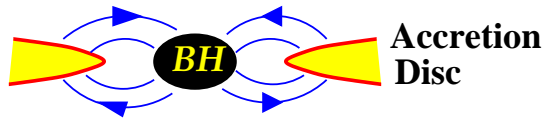,height=0.75in}}
\caption{Magnetic field lines, anchored in the disk, which thread
the black hole, coupling the disk rotation to that of the black hole.
The more rapidly rotating black hole torques up the accretion disk
delivering rotational energy into it. This energy is changed to heat
by the viscosity, powering the hypernova explosion.}
\label{mbreak}
\end{figure}

Blandford \& Spruit\cite{Blandford00} and Van Putten\cite{putten01}
have shown that important dissipation into
the disk comes through magnetic field lines coupling the disk to the
black hole rotation. As shown in Fig.~\ref{mbreak} these lines,
anchored in the inner disk, thread the black hole.

The more rapidly rotating black hole will provide torques, along its
rotation axis, which spin up the inner accretion disk, in which the closed
magnetic field lines are anchored. With increasing centrifugal force
the material in the inner disk will move outwards, cutting down the
accretion. Angular momentum is then advected outwards, so that the
matter can drift back inwards. It then delivers more matter to
the black hole and is flung outwards again. The situation is like that of
a ball in a roulette wheel (R.D. Blandford, private communication).
First of all it is flung outwards and then drifts slowly inwards.
When it hits the hub it is again thrown outwards.
The viscous inflow time for the fluctuations is easily estimated to be
   \be
   \tau_{visc}\sim \Omega^{-1}_{disk}
   \left(\frac{r}{H}\right)^2 \alpha_{vis}^{-1}
   \label{visc}
   \ee
where $H$ is the height of the disk at radius $r$, and $\alpha_{vis}$
is the usual $\alpha$-parameterization of the viscosity. We choose
$\alpha_{vis}\sim 0.1$, $r/H\sim 10$ for a thin disk and then arrive at
$\tau_{\rm visc}\sim 0.1$\,s. We therefore expect variability on all time scales
between the Kepler time (sub-millisecond) and the viscous time, which may
explain the very erratic light curves of many GRBs.

We suggest that the GRB can be powered by $\dot E_{\rm GRB}$ and a
Type Ibc supernova explosion by $\dot E_\SN$ where $\dot E_{\SN}$
is the power delivered through dissipation into the disk. To the
extent that the number of closed field lines coupling disk and
black hole is equal to the number of open field lines threading
the latter, the two energies will be equal. In the spectacular
case of GRB 980326~\cite{Bloom99C}, the GRB
lasts about $5$\,s, which we take to be the time that the central
engine operates. We shall show that up to $\sim 10^{53}$\,erg is
available to be delivered into the GRB and into the accretion
disk, the latter helping to power the supernova (SN) explosion.
This is more energy than needed and we suggest that
injection of energy into the disk shuts off the central engine by
blowing up the disk and thus removing the magnetic field needed for the
energy extraction from the black hole. If
the magnetic field is high enough the energy will be delivered in
a short time, and the quick removal of the disk will leave the black hole
still spinning quite rapidly.

\subsection{Energetics of GRBs}

The maximum energy that can be extracted from the BZ mechanism~\cite{Lee99A} is
   \be
   (E_{BZ})_{max} \simeq 0.09\ M_\BH c^2.
   \ee
This is 31\% of the black hole rotational energy, the remainder going
toward increasing the entropy of the black hole. We have chosen here
(see preceding footnote)
   \be
   \epsilon_\Omega
   =\frac{\Omega_{disk}}{\Omega_{H}}=0.5
   \label{eqeps}
   \ee
which maximizes the BZ power.

For a $7\msun$ black hole, such as that found in Nova Sco 1994
(GRO\,J1655$-$40),
   \be
   E_{max} \simeq 1.1\times 10^{54}\ {\rm erg}.
   \label{eq5}\ee
We estimate below that the energy available in a typical case will
be an order of magnitude less than this.
Without collimation, the estimated gamma-ray energy in GRB\,990123 is
about $4.5\times 10^{54}$\,erg~\cite{Andersen99}.
The BZ scenario entails substantial beaming, so this energy should be multiplied
by $d\Omega/4\pi$, which may be a small factor $\sim 10^{-2}$.

The BZ power can be delivered at a maximum rate~\cite{Lee99A} of
    \be
    P_{\rm BZ} =6.7\times 10^{50}\left(\frac{B}{10^{15}{\rm G}}\right)^2
    \left(\frac{M_\BH}{\msun}\right)^2\ {\rm erg\ s^{-1}},
    \label{eq7}
    \ee
so that high magnetic fields are necessary for rapid delivery.

The above concerns the maximum energy output into the jet and the disk.
The real energy available in black-hole spin in any given case, and the
efficiency with which it can be extracted, depend on the rotation
frequency of the newly formed black hole and the disk or torus around it.
The state of the accretion disk around the newly formed black hole, and
the angular momentum of the black hole, are somewhat uncertain. However,
the conditions should be bracketed between a purely Keplerian, thin disk
(if neutrino cooling is efficient) and a thick, non-cooling hypercritical
advection-dominated accretion disk (HADAF)~\cite{Brown00}. 
Let us examine the result for the Keplerian case.  In terms of
   \be
   \tilde a=\frac{Jc}{M^2 G}
   \ee
we find the rotational energy of a black hole to be
   \be \label{erotfull}
   E_{rot} = f(\tilde a) M c^2,
   \ee
where
   \be
   f(\tilde a)=1-\sqrt{\frac 12 (1+\sqrt{1-{\tilde a}^2})}.
   \label{eqfa}
   \ee
For a maximally rotating black hole one has $\tilde{a}=1$\footnote{
   As an aside, we note a nice mnemonic: if we define a velocity $v$
   from the black-hole angular momentum by $J=MR_\Sch v$, so that $v$
   carries the quasi-interpretation of a rotation velocity at the
   horizon, then $\tilde{a}=2v/c$. A maximal Kerr hole, which has
   $R_{\rm event}=R_\Sch/2$, thus has $v=c$. For $\tilde{a}\lsim0.5$,
   the rotation energy is well approximated by the easy-to-remember
   expression $E_{\rm rot} = \frac{1}{2}Mv^2$.}.

We begin with a neutron star in the middle of a Keplerian
accretion disk, and let it 
accrete enough matter to send it into a black hole.

In matter free regions the last stable orbit of a particle around
a black hole in Schwarzschild geometry is
    \be
    r_{\rm lso}=3 R_{\Sch} =6\frac{GM}{c^2}.
    \ee
This is the marginally stable orbit $r_{\rm ms}$.
However, under conditions of hypercritical accretion,
the pressure and energy profiles are changed and it is better
to use~\cite{Abramowicz88}
    \be
    r_{\rm lso}\gsim 2 R_{\Sch}.
    \ee
With the equal sign we have the marginally bound orbit $r_{\rm mb}$.
With high rates of accretion we expect this to be a  good approximation
to $r_{\rm lso}$.
The accretion disk can be taken to extend down to the last stable
orbit~\cite{grb2000}.

We take the angular velocity to be Keplerian, so that
at the radius of $2 R_{\Sch}$
    \be
    v^2 = \frac{GM}{2 R_{\Sch}}=\frac{c^2}{4},
    \ee
or $v=c/2$. The specific angular momentum is then
    \be
    l \ge 2 R_{\Sch} v = 2\frac{GM}{c}
    \label{eq16}\ee
which in Kerr geometry indicates $\tilde a\sim 1$. Had we taken one
of the slowest-rotating disk flows that are possible, the advection-dominated
or HADAF case~\cite{Narayan94,Brown00}, which
has $\Omega^2=2\Omega_K^2/7$, we would have arrived at $\tilde a\sim 0.54$,
so the Kerr parameter will always be high.
Later in Sec.~\ref{rotate}, we discuss that the Kerr parameter
for the SXTs with orbital period shorter than or similar to
0.5 days, at the time of black hole formation,
is very large, close to the maximum rotation.

Further accretion will add angular momentum to the black hole at a rate
determined by the angular velocity of the inner disk.
The material accreting into the
black hole is released by the disk at $r_{\rm lso}$, where the angular
momentum delivered to the black hole is determined. This angular
momentum is, however, delivered into the black hole at the event
horizon $R_{\Sch}$, with velocity at least double that at which it
is released by the disk, since the lever arm at the event horizon
is only half of that at $R_{\Sch}$, and angular momentum is
conserved. With more rapid
rotation involving movement towards a Kerr geometry where the
event horizon and last stable orbit coincide at
   \be
   r_{\rm lso}=R_{\rm event}=\frac{GM}{c^2}.
   \ee
Although we must switch over to a Kerr geometry for quantitative results,
we see that $\tilde a$ will not be far from its maximum value of unity.
Again, for the lower angular-momentum case of a HADAF, the expected
black-hole spin is not much less.

Van Putten \cite{putten2000,putten01} has listed many channels
through which the rotational energy of the black hole can be dissipated.

We note that in the scenario we developed here, the black hole rotational
energy will depend only on the binary period just before black hole formation,
which, in turn, is determined chiefly by companion mass~\cite{Lee2002}.
However, the power, the rate at which the energy is delivered,
is proportional to $B^2$, which will vary from star to star.
If $B^2$ is low, so that the central engine must work for a long time,
then neutrino losses will be somewhat increased.

Finally we note that an extensive numerical calculation by 
Koide et al.\cite{Koide02} supports in a general way
that the rotational energy can be extracted from black holes as
in Blandford-Znajek, although the energy comes out in somewhat
different forms.

\subsection{Hypernova explosion}
\label{sec2.2}

In Brown et al.\cite{grb2000} the Blandford-Znajek mechanism was described
in a schematic way; first of all a rapidly rotating black hole evolves
in the core of the He star in a binary. The He star had been set into
rotation as the H envelope of the massive black hole progenitor was
taken off by the companion star in common envelope evolution and further
spun up by tidal interactions as described by Lee et al.\cite{Lee2002}. 
The rotational energy of the black hole was used both to power the
GRB and the hypernova explosion, the latter in the
accretion disk resulting from the inner part of the remnant helium star.

The way in which the rotating black hole powers the hypernova explosion
was developed by Van Putten\cite{putten99,putten01}. The black hole rotates much
faster than the accretion disk, so the closed magnetic field lines
threading the black hole at one end and frozen in the matter of the
accretion disk at the other end torque the latter up, delivering angular
momentum and energy into the latter. Li\cite{Li2000b} has shown that
the efficiency for the black hole to deliver energy into the accretion 
disk is greater than that into the GRB. As much as 
$\sim 10^{52}$~ergs can be provided for the hypernova explosion, if the 
black hole rotates sufficiently fast.

Brown et al.\cite{grb2000} showed that the outer part of the spinning 
He star can be supported by centrifugal force for a viscous time scale,
Eq.~(\ref{visc}) which will be much longer than the 
$\tau_{\rm visc}$ for the inner disk because of the lower $\Omega$
and smaller $\alpha$ far out in the He star.
(The $\alpha$ will be smaller because of the lower magnetic field
B\cite{brandenburg96}.)
For thin disks this can be two orders of magnitude longer
than the dynamical time which we discuss later.
By this time the hypernova explosion is assumed to reach the matter
and expel it. Once the energy from the black hole is converted into
heat by viscosity, the hypernova explosion is triggered in a relatively
short time. Given the $10^{52}$ ergs energy estimated in 
Brown et al.\cite{grb2000},
and mass $2\times 10^{34}$ g, we find a velocity
   \be
   v=\sqrt{\frac{2E}{M}}=10^9\; {\rm cm\ s^{-1}} .
   \ee
With a He-star radius $\lsim 1 \rsun$ it takes the shock $\sim 50$ s,
roughly the viscous time scale, to go through the outer He star, although
the subsequent expansion develops over a much longer time.

The empirical relationship between preexplosion period, which was 
reconstructed through evolutionary arguments, and black hole mass
was adduced by Lee et al.\cite{Lee2002} to support this argument. The close
relationship with Woosley's failed supernova model for GRBs~\cite{Woosley93B}
was noted.

Because with black hole formation the He star fell into an accretion disk
this model is clearly aspherical.
Podsiadlowski et al.\cite{Pods02} investigated both spherical and aspherical
models for matter deposition onto the F-star companion.
They found that with the $16\msun$ He star non-spherical models they
calculate, all produce either an unacceptable overabundance of
O and Mg or an unacceptable underabundance of S and Si, depending on
where the cut-off below which matter can be mixed into the ejecta occurs.
However, they find for spherical models that 10 to $16\msun$ He cores
are the most probable, consistent with the $11\msun$ He star found by 
Lee et al.\cite{Lee2002}. O and Mg are formed in quiescent burning in stars
and their abundance in the presupernova model is roughly linear with
ZAMS mass, and decreasing it from $40\msun$ to $30\msun$ will reduce
these elements. The amounts of Si and S depend sensitively on the energy
of the explosion. We believe that aspherical calculations should be
made in the $30\msun$ region of ZAMS masses.

We believe Nova Scorpii should be evolved like the
other SXTs, it being exceptional because of its low preexplosion period,
and its loss of nearly half of its mass in the explosion forming
the black hole.  Its system velocity
is large in a Blaauw-Boersma type kick, because the companion is
relatively massive, $\sim 2\msun$ at the time of explosion as we
detail later, so the center of gravity is somewhat away from the black hole
where the mass is lost.

The SXTs with low-mass main sequence companion do not have large system
velocities~\cite{Nelemans99}, at least not radial ones, which would
be seen. Their black hole evolution should not be influenced appreciably
by the companion star. On the other hand, the Blaauw-Boersma kick
is roughly proportional to the companion mass and the mass loss being at the
position of the black hole, so that the system velocity depends on the 
difference in position of the black hole and center of gravity of the
system.

We suggest that the neutron star kick velocity may be suppressed by the
rapid rotation of the system in which it is born.
A possible model for the neutron star kick velocity in supernovae is
that it arises from the stochastic sowing of convection cells necessary
for the supernova explosion. Because of the randomness, the number of cells
in one hemisphere can be different from the number in the other, and this
can power the kick velocity in a jet-like fashion.
In the accretion disk preceeding neutron star formation, the Bernoulli number
is positive and convection carries matter out at angles $\theta\sim 30^\circ
-45^\circ$~\cite{MacFadyen99}. Whereas the MacFadyen-Woosley\cite{MacFadyen99}
treatment is numerical,
the underlying accretion disk equations are symmetrical in polar angle and do
not involve the angle of rotation about the axis, so one would not expect
any  substantial inbalance in the two hemispheres as in the supernova problem.

\subsection{Previous models}

\subsubsection{Collapsar}

Woosley\cite{Woosley93B}, and MacFadyen \& Woosley\cite{MacFadyen99} 
suggested the Collapsar model as a source of GRBs.
In this model the center of a rotating
Wolf-Rayet star evolves into a black hole, the outer part being held
out by centrifugal force. The latter evolves into an accretion disk and
then by hypercritical accretion spins the black hole up. 

\begin{figure}
\centerline{\epsfig{file=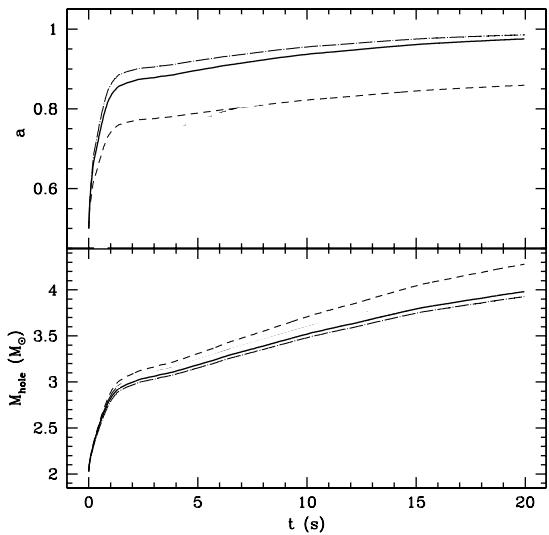,height=3.0in}}
\caption{Time evolution of BH mass and angular momentum taken
from Fig. 19 of MacFadyen \& Woosley\protect\cite{MacFadyen99}.
The upper panel shows the increase in the Kerr parameter for various
models for the disk interior to the inner boundary at 50 km.
``Thin" (dash-dot), neutrino-dominated (thick solid) and advection
dominated (short dash) models are shown for initial Kerr parameter
$\tilde a_{init}=0.5$.
The lower panel shows the growth of the gravitational mass of the
black hole. The short-dashed line shows the growth in baryonic mass of
the black hole since for a pure advective model no energy escapes the
inner disk.
}
\label{woosley}
\end{figure}

Our schematic model has the advantage over numerical calculations
that one can see analytically how the scenario changes with change
in parameters or assumptions. However, our model is useful only if
it reproduces faithfully the results of more complete calculations
which involve other effects and much more detail than we include.
We here make comparison with Fig.19 of MacFadyen \& Woosley\cite{MacFadyen99}.
Accretion rates, etc., can be read off from their figure
which we reproduce as our Fig.\ref{woosley}. MacFadyen \& Woosley
prefer $\tilde{a}_{initial}=0.5$ (We have removed their curve for
$\tilde{a}_{initial}=0$). This is a reasonable value if the black hole
forms from a contracting proto-neutron star near breakup.
MacFadyen \& Woosley find that $\tilde{a}_{initial}=0.5$ is more
consistent with the angular momentum assumed for the mantle than
$\tilde{a}_{initial}=0$. (They take the initial black hole to have mass
$2\msun$; we choose the Brown \& Bethe\cite{Brown94} mass of $1.5\msun$.)
We confirm this in the next section.

After 5 seconds (the duration of GRB\,980326) the MacFadyen \&
Woosley black hole mass is $\sim 3.2\msun$ and their Kerr
parameter $\tilde{a}\sim 0.8$, which gives $f(\tilde a)$ of our
Eq.(\ref{eqfa}) of 0.11. With these parameters
we find $E=2\times 10^{53}$ erg,
available for the GRB and SN explosion.

One can imagine that continuation of the MacFadyen \& Woosley
curve for $M_{BH}(\msun)$ would ultimately give something like our
$\sim 7\msun$, but the final black hole mass may not be relevant
for our considerations. This is because more than enough energy is
available to power the supernova in the first 5 seconds; as the
disk is disrupted, the magnetic fields supported by it will also
disappear, which turns off the Blandford-Znajek mechanism.

Power is delivered at the rate given by Eq.(\ref{eq7}). Taking a 
black hole mass relevant here, $\sim 3.2\msun$, we require a 
field strength of $\sim 5.8\times 10^{15}$\,G 
in order for our estimated energy ($4\times 10^{52}$ erg) to be
delivered in 5\,s (the duration of GRB\,980326).
For such a relatively short burst, we see that the required field 
is quite large, but it is still not excessive if we bear in mind that
magnetic fields of $\sim10^{15}$\,G have already been observed in
magnetars\cite{Kouveliotou98,Kouveliotou99A}.
Since in our scenario we have many more progenitors than 
there are GRBs, we suggest that the necessary fields are obtained only in
a fraction of all potential progenitors.

Thus we have an extremely simple scenario for powering a GRB and the
concomitant SN explosion in the black hole transients, which we will
discuss in Sec.\ref{rotate}. After the first second the newly
evolved black hole has $\sim 10^{53}$ erg of rotational energy
available to power these. The time scale for delivery of this
energy depends (inversely quadratically) on the magnitude of the
magnetic field in the neighborhood of the black hole, essentially
that on the inner accretion disk. The developing supernova
explosion disrupts the accretion disk; this removes the magnetic fields
anchored in the disk, and self-limits the energy the B-Z mechanism can deliver.

MacFadyen \&
Woosley point out that ``If the helium core is braked by a magnetic field
prior to the supernova explosion to the extent described by Spruit \&
Phinney\cite{Spruit98} then our model will not work for single
stars." Spruit \& Phinney argue that magnetic fields maintained by
differential rotation between the core and envelope of the star will
keep the whole star in a state of approximately uniform rotation until
10 years before its collapse.  As noted in the last section, with the
extremely high magnetic fields we need the viscosity would be expected
to be exceptionally high, making the Spruit \& Phinney scenario probable.
Livio \& Pringle\cite{Livio98} have commented that one finds
evidence in novae that the coupling between layers of the star by magnetic
fields may be greatly suppressed relative to what Spruit \& Phinney assumed.
However, we note that  even with this suppressed coupling, they find
pulsar periods from core collapse supernovae no shorter than 0.1\,s.
Independent evidence for the fact that stellar cores mostly rotate no faster
than this comes from the study of supernova remnants: 
Bhattacharya\cite{Bhatta90,Bhatta91}
concludes that the absence of bright, pulsar-powered plerions in most SNRs
indicates that typically pulsar spin periods at birth are no shorter than
0.03--0.05\,s.
Translated to our black holes, such spin periods would imply
$\tilde{a}\lsim0.01$, quite insufficient to power a GRB.

\subsubsection{Coalescing low-mass black holes and helium stars}
\label{coalesc}

Fryer \& Woosley\cite{Fryer98} suggested the scenario of a black hole spiraling
into a helium star. This is an efficient way to spin up the
black hole.

Bethe \& Brown\cite{Bethe98} evolved low-mass black holes with helium star
companion, as well as binaries of compact objects. In a total available
range of binary separation $0.04 < a_{13}<4$, low-mass black-hole,
neutron-star binaries were formed when $0.5<a_{13}<1.4$ where
$a_{13}$ is the initial binary separation in units of $10^{13}$ cm.
The low-mass black hole coalesces with the helium star in the range
$0.04<a_{13}<0.5$. Binaries were distributed logarithmically in $a$.
Thus, coalescences are more common than low-mass black-hole,
neutron-star binaries by a factor of $\ln(0.5/0.04)/\ln(1.9/0.5)=1.9$

In Bethe \& Brown\cite{Bethe98}, the He-star, compact-object binary was
disrupted $\sim 50\%$ of the time by the He-star explosion.
This does not apply to the coalescence. Thus, the rate of
low-mass black-hole, He-star mergers is 3.8 times the formation
rate of low-mass black-hole, neutron-star binaries, or
   \be
   R=3.8\times 10^{-4}\ {\rm yr}^{-1}
   \ee
in the Galaxy. The estimated empirical rate of GRBs, with a factor
of 100 for beaming, is $10^{-5}$ yr$^{-1}$ in the Galaxy\cite{Brown99B}. 
Thus, the number of progenitors is more than adequate.

In Bethe \& Brown\cite{Bethe98}
the typical black hole mass was $\sim 2.4
M_\odot$, somewhat more massive than their maximum assumed neutron
star mass of $1.5 M_\odot$. As it enters the helium star companion
an accretion disk is soon set up and the accretion scenario will
follow that described above, with rotating black holes of various
masses formed.

The problem with this scenario is, however, that usually a
substantial hydrogen
envelope will be left around the black hole, helium star system.
The reason the black hole coalesces with the He star is that its drop
in gravitational binding is insufficient to expel the hydrogen envelope.
If an appreciable envelope is left, it would be difficult for a jet to
return a highly relativistic $\Gamma$ while plowing through the overlying
matter~\cite{MacFadyen99}.
A layer of surface hydrogen is allowed
as long as its radius is not too large, but even a modest total amount
of hydrogen will interfere with a jet. The supernova, however, might still
look like an exceptionally powerful type Ibc event.
Our estimated number of coalescing
low-mass black-hole, He star systems is so large that some interesting
SN explosions and perhaps even GRBs should nonetheless survive.

\section{SOFT X-RAY TRANSIENTS: OBSERVATIONS}
\label{sxts}
\subsection{Hypernova explosion in Nova Scorpii}

Far and away the most interesting binary to date is Nova Scorpii
although GRS 1915$+$105 is now growing in importance
since the companion star has been found (see later).
Nova Sco 1994 is a black hole transient X-ray source. It consists
of a black hole of $\sim 5.4\msun$ and a subgiant of
current mass of about $1.45\msun$.
Their separation is $15.2\rsun$.
Israelian et al. (1999) have analyzed the spectrum of the subgiant
and have found that the $\alpha$-particle nuclei O, Mg, Si and S
have abundances 6 to 10 times the solar value (see Table~\ref{tabNSco}). 
This indicates that
the subgiant has been enriched by the ejecta from a supernova
explosion; specifically, that some of the ejecta of the supernova
which preceded the present Nova Sco (a long time ago) were
intercepted by star B, the present subgiant. 
Israelian et al.\cite{Israelian99}
estimate an age since accretion started from the assumption that
enrichment has only affected the outer layers of the star. We here
reconsider this: the time that passed since the explosion of the 
progenitor of the black hole is roughly the main-sequence lifetime
of the present subgiant companion, which given its mass of $\sim$2$\msun$
will be about 1\,Gyr. This is so much longer than any plausible mixing
time in the companion that the captured supernova ejecta must by now be 
uniformly mixed into the bulk of the companion. This rather increases
the amount of ejecta that we require the companion to have captured.
(Note that the accretion rate in this binary is rather less than expected
from a subgiant donor, though the orbital period leaves no doubt that
the donor is more extended than a main-sequence star\cite{Regos98}. 
It is conceivable that the high metal abundance
has resulted in a highly non-standard evolution of this star, in which case
one might have to reconsider its age.)

\begin{table}
\caption{Abundances in the secondary of Nova Scorpii\protect\cite{Israelian99}.
[Xi/H] are the logarithmic abundances relative to solar abundances.}
\vskip 3mm
\begin{center}
\begin{tabular}{lccccccc} \hline
       &   N    &  O   & Mg   &  Si  &   S  & Ti   & Fe \\ \hline
[Xi/H] & 0.45   & 1.00 & 0.90 & 0.90 & 0.75 & 0.90 & 0.10 \\
error  & 0.50   & 0.30 & 0.40 & 0.30 & 0.20 & 0.40 & 0.20 \\ \hline
\end{tabular}
\end{center}
\label{tabNSco}
\end{table}

\begin{figure}
\epsfig{file=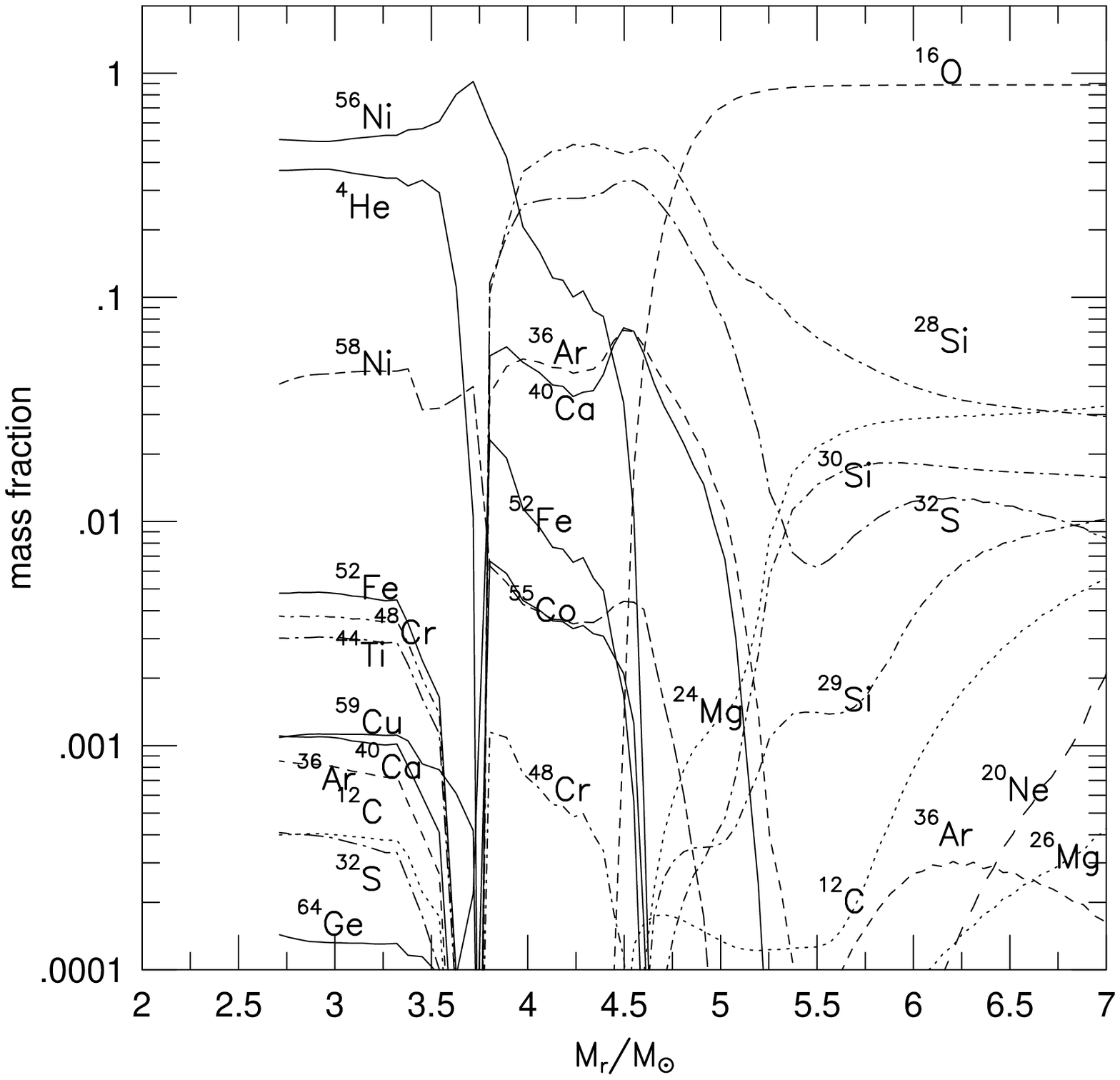,width=0.48\textwidth}\hfill\epsfig{file=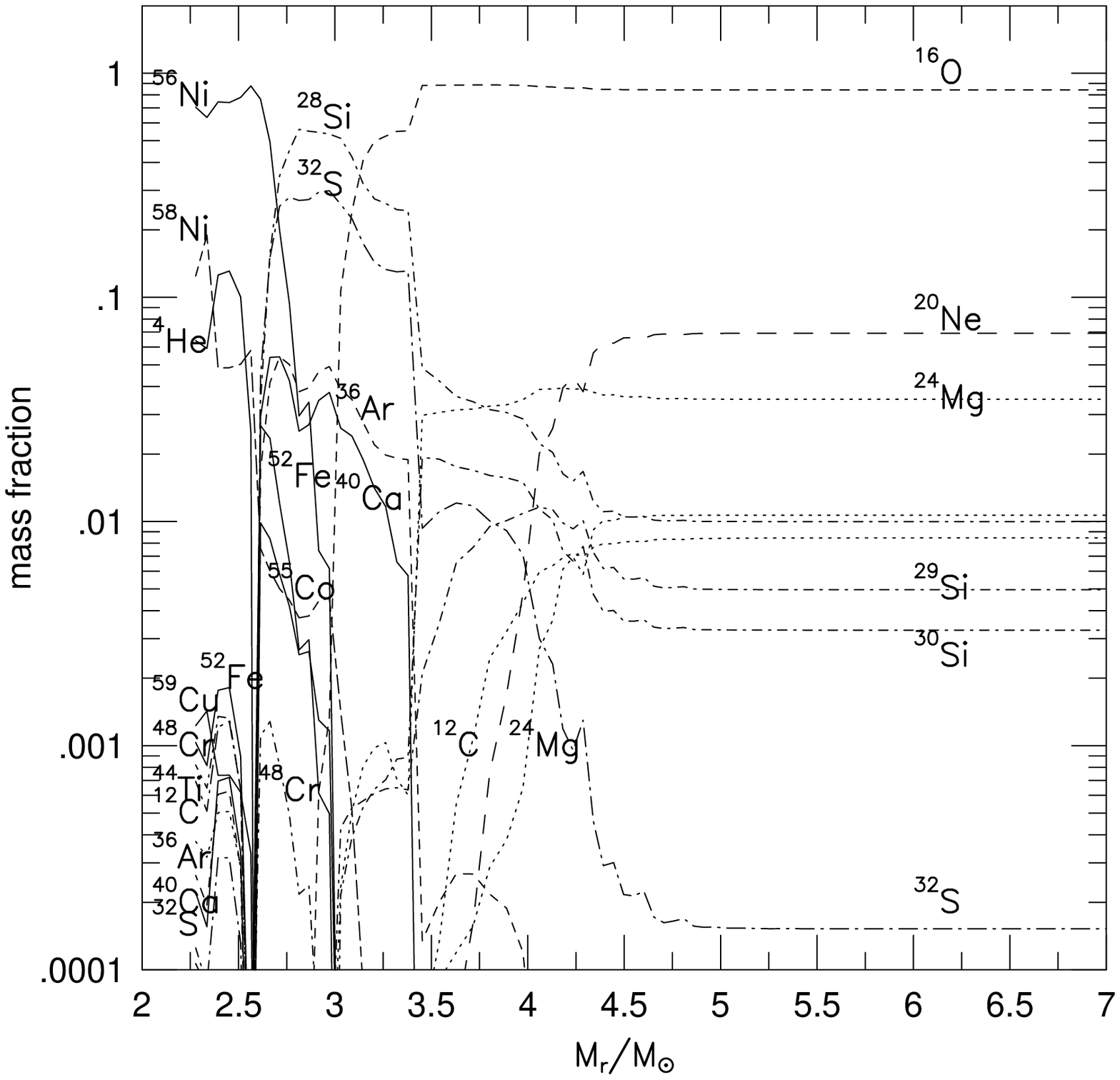,width=0.48\textwidth}
\caption{
The isotopic composition of ejecta of the hypernova 
($E_{\rm K}=3\times10^{52}$\,erg; left) and the normal supernova 
($E_{\rm K}=1\times10^{51}$\,erg; right)
for a $16\msun$ He star, from Nomoto et~al.\protect\cite{Nomoto00}. 
Note the much higher sulphur abundance in the hypernova.
}\label{fig1i}
\end{figure}

The presence of large amounts of S is particularly significant.
Nomoto et al.\cite{Nomoto00} have calculated the composition of a
hypernova from an $11\msun$ CO core, see Fig.~\ref{fig1i}. This
shows substantial abundance of S in the ejecta. Ordinary supernovae
produce little of this element, as shown by the results of Nomoto
et al.\cite{Nomoto00} in Fig.~\ref{fig1i}.
The large amount of S, as well as O, Mg and Si we
consider the strongest argument for considering Nova Sco 1994 as a
relic of a hypernova, and for our model, generally.

Fig.~\ref{fig1i} also shows that $^{56}$Ni and $^{52}$Fe are
confined to the inner part of the core, below the mass cut.
However, as known from SN 1987A, $^{56}$Ni (which decays into
$^{52}$Fe) gets out through Rayleigh-Taylor instabilities.
Maeda et al.\cite{Maeda02} as described by Nomoto et al.\cite{nomoto01}
find that in the aspherical (pancake-type) explosion appropriate
for the hypernova, the $^{56}$Ni goes out  preferentially 
perpendicular to the pancake. Thus, the low amount of Fe
deposited on the F-star companion in Nova Scorpii has a natural
explanation because we see the binary edge on. On the other hand,
not much of a GRB was seen in 1998bw, which is interpreted in terms
of us seeing this binary sideways. We would, however, be expected
to see a lot of $^{56}$Ni. We are grateful to Garik Israelian
for pointing this out to us.

The solar abundance of oxygen is about 0.01 by mass, so with the abundance
in the F star being 10 times solar, and oxygen uniformly mixed, we expect
$0.1\times2.5=0.25\msun$ of oxygen to have been deposited on the companion.
[Si/O] is 0.09 by mass in the Sun, and
[S/O] is 0.05, so since the over-abundances of all three elements are similar
we expect those ratios to hold here, giving about 0.02$\msun$ of captured Si
and 0.01$\msun$ of captured S. We therefore need a layer of stellar ejecta
to have been captured which has twice as much Si as S, at the same
time as having about 10 times more O. From Fig.~\ref{fig1i}, we see
that this occurs nowhere in a normal
supernova, but does happen in the hypernova model\cite{Nomoto00}
at mass cuts of 6$\msun$ or more. This agrees very nicely with the notion
that a hypernova took place in this system, and that the inner 7$\msun$ or
so went into a black hole.
In Sec.~\ref{other} the chemical deposition will be discussed
in more detail.

A further piece of evidence that may link Nova Sco 1994 to our GRB/hypernova
scenario are the
indications that the black hole in this binary
is spinning rapidly.  Zhang et al.\cite{Zhang97} argue from
the strength of the ultra-soft X--ray component that the black hole
is spinning near the maximum rate for a Kerr black hole.  However,
studies by Sobczak et al.\cite{Sobczak99} show that it must be
spinning with less than 70\% maximum.  Gruzinov\cite{Gruzinov99}
finds the inferred black hole spin to be about 60\% of maximal from
the 300 Hz QPO. Strohmayer\cite{Strohmayer01} also found the 450 Hz QPO, 
which also
support the rapidly rotating black hole in Nova Sco.
Our estimates of the last section indicate that enough
rotational energy will be left in the black hole so that it will still
be rapidly spinning\cite{Lee2002}.

We have already mentioned the unusually high space  velocity of
$-106 \pm 19$ km s$^{-1}$. 
Its origin was first discussed by 
Brandt et~al.\cite{Brandt95}, who concluded that significant mass must have been
lost in the formation of the black hole in order to explain this high
space velocity:
it is not likely
to acquire a substantial velocity in its own original frame of
reference, partly because of the large mass of the black hole. But
the mass lost in supernova is ejected from a moving object and thus
carries net momentum. Therefore, momentum conservation demands that
the center of mass of the binary acquire a velocity;
this is the Blaauw--Boersma kick\cite{Blaauw61,Boersma61}. 
Note that the F-star companion mass is relatively large among the black-hole
transient sources, so the center of mass is somewhat away from the black hole
and one would expect a large kick.
Nelemans et al.\cite{Nelemans99} estimate
the mass loss in this kick to be $5-10\msun$.

In view of the above, we consider it well established that 
the black hole in Nova Sco 1994 is the relic of a hypernova. 
We believe it highly likely that some of the
other black-holes in the transient X-ray sources are also hypernova remnants. 
Later in Sec.~\ref{rotate} we will show that the SXTs with
short orbital period at the time of black hole formation are 
good candidates for fossil remnants of GRBs.

\subsection{An empirical correlation between orbital period and black-hole mass
         \label{porb}}

\begin{figure}
\centerline{\epsfig{file=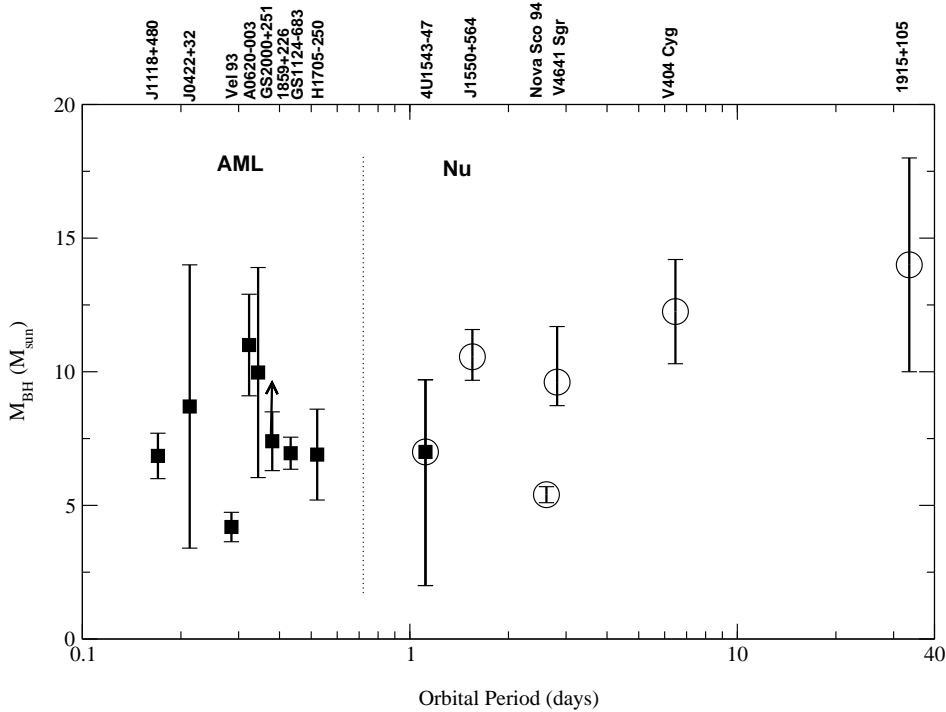,height=4in}}
\caption{Black hole mass as a function of present orbital period of 14 SXTs. 
Note that the orbital period is on a logarithmic scale. 
SXTs with subgiant or giant companions
are indicated by big open circles (denoted as ``Nu" for nuclear evolution). 
Filled squares indicate 
SXTs with main-sequence companions 
(denoted as ``AML" for angular momentum loss).
The vertical dotted line is drawn to indicate the possible
existence of different classes according to evolutionary path of the
binary, as discussed in section~\ref{evol}.
4U 1543$-$47 is marked with both symbols, since we believe it to
be right on the borderline between main-sequence and evolved; for the
purpose of modeling, it can be treated as evolved.
}
\label{FIG1}
\end{figure}

\begin{figure}
\centerline{\epsfig{file=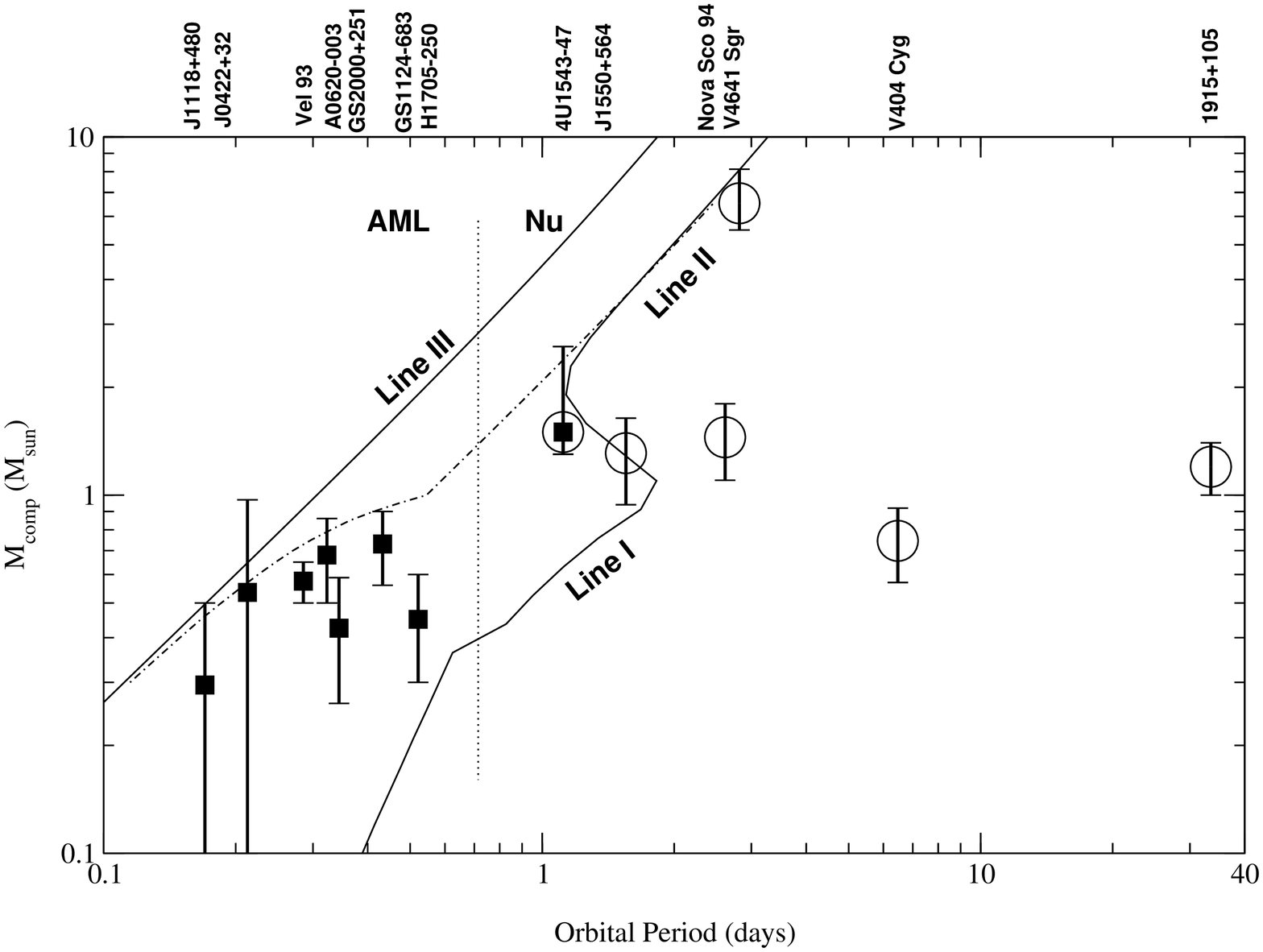,height=4.0in}}
\caption[]{Companion mass as a function of present orbital period of 13 SXTs. 
XTE 1859$+$226 is not included because the companion mass
is not well determined\cite{Filippenko01}.
Symbols of SXTs are the same as in Fig.~\ref{FIG1}. Line III indicates the
orbital period for which a companion of that mass fills its Roche lobe on
the ZAMS. No system can exist above and to the left of this line for 
a significant duration. Lines I and II are the upper period limit for systems
that can come into contact while the donor is on the main sequence. For
high masses (line II) this limit is set by the period where the evolution
time of the companion is too short to allow the orbit to shrink significantly
before it leaves the main sequence.  For low masses
(line I), where the donor never evolves off the main sequence within a 
Hubble time, the limit is set by period for which the shrinking time scale
of the orbit equals the Hubble time. The dot-dashed line indicates the point
where a system that starts its life on lines I/II comes into Roche contact.
For very low masses, this equals line III, because the donor never 
moves significantly away from its ZAMS radius, whereas for very high masses
it equals line II, because the orbit cannot shrink before the companion
evolves off the main sequence. At intermediate masses, the companion expands
somewhat while the orbit shrinks, and fills its Roche lobe at a larger
period than line III. Systems that become SXTs with main-sequence donors
within a Hubble time must start between line III and line I/II. At the start
of mass transfer, they must lie in the narrow strip between line III and
the dot-dashed line.
}
\label{FIG2}
\end{figure}

We have collected data from the literature on black-hole binaries in
our Galaxy. In Table~\ref{tab1} we collect data of those in which the 
mass function
is known, and some manner of mass estimate for both the black hole and 
companion can be given. 
In Table~\ref{tab2} we list the properties of two key systems
in more detail. In Fig.~\ref{FIG1} we show the masses of the black
holes as a function of orbital period. While the ranges of black-hole
masses for main-sequence and evolved systems overlap, the latter tend to
have higher masses; the exception is Nova Scorpii 1994, which we shall
see later is a natural but rare case of the general evolution scenario
that we describe in this paper. In Fig.~\ref{FIG2}, we show the 
donor masses as a function of orbital period. They show a more obvious
trend of more massive donors in evolved systems. As we shall see, this
is a natural consequence of the fact that only evolved systems can come
into Roche contact in wide binaries, and more massive donors are more
likely to come into contact via nuclear evolution. (The various curves
are explained in section~\ref{evol}.)

\begin{table}
\caption{Parameters of black hole binaries in our Galaxy with 
measured mass functions.
Binaries are listed in order of increasing orbital period.
All systems except Cyg~X-1 (steady X-ray source) are SXTs.
         XN indicates X-ray Nova.
Earlier observations (Greene et al. 2001) gave
the black hole mass in Nova Scorpii as $6.3\pm 0.5\msun$
New analyses of the light curve by Beer \& Podsiadlowski (2001)
give a somewhat smaller mass $5.4\pm 0.3 \msun$ and $1.45\pm 0.35\msun$
for the companion. 
References: 
a) McClintock et al. 2001\protect\cite{McClintock01},
b) Wagner et al. 2001\protect\cite{Wagner01},
c) Bailyn et al. 1998\protect\cite{Bailyn98},
d) Filippenko et al. 1999\protect\cite{Filippenko99},
e) Gelino et al. 2001a\protect\cite{Gelino01A},
f) Harlaftis et al. 1996\protect\cite{Harlaftis96},
g) Filippenko \& Chornock 2001\protect\cite{Filippenko01}, 
h) Gelino et al. 2001b\protect\cite{Gelino01B},
i) Orosz et al. 1998\protect\cite{Orosz98},
j) Orosz  2002\protect\cite{Orosz02A}, 
k) Orosz et al. 2002\protect\cite{Orosz02B}, 
l) Beer \& Podsiadlowski 2001\protect\cite{Beer02},
m) Orosz et al. 2001\protect\cite{Orosz01},
n) Herrero et al. 1995\protect\cite{Herrero95},
o) Shahbaz et al. 1994\protect\cite{Shahbaz94},
p) Shahbaz et al. 1996\protect\cite{Shahbaz96},
q) Greiner et al. 2001\protect\cite{Greiner01}. 
}
\vskip 3mm
\begin{center}
{\footnotesize
\def\arraystretch{0.9}
\newcommand{\ti}[1]{{\footnotesize #1}}
\begin{tabular}{@{}lccccl@{}}\hline
                    &   compan.         & $P_{orb}$             & $f(M_{X})$ & $M_{opt}$              &      \\ 
X-ray               &   type            & (day)                 &  ($\msun$) & ($\msun$)              & Ref. \\
names               &   $K_{opt}$       &      $d$              &  i         & $M_{BH}$               & \\ 
                    &   (km s$^{-1}$)   & (\kpc)                &  (degree)  & ($\msun$)              & \\
\hline
XTE J1118$+$480       & K7V-M0V    & 0.169930(4)           & 6.1(3)     &  0.09--0.5             & a,b \\
\ti{KV Ursae Majoris} & 701(10)    & 1.9(4)                & 81(2)      &  6.0--7.7              & \\ \hline
XN Persei 92          &  M0 V      & 0.2127(7)             & 1.15--1.27 & 0.10-0.97              & c\\
\ti{GRO\,J0422$+$32}  &  380.6(65) &                       & 28--45     & 3.4--14.0              & \\ \hline
XN Vel 93             & K6-M0      & 0.2852                & 3.05--3.29 & 0.50--0.65             & d \\
\ti{MM Velorum}       &  475.4(59) &                       & $\sim$ 78  &  3.64--4.74            & \\ \hline
XN Mon 75             &  K4 V      & 0.3230                & 2.83-2.99  & 0.68(18)               & c,e\\
\ti{A\,0620$-$003}    & 433(3)     &  1.164(114)           & 40.75(300) & 11.0(19)               & \\ \hline
XN Vul 88             &  K5 V      & 0.3441                &  5.01(12)  & 0.26--0.59             & c,f \\
\ti{GS\,2000$+$251}   &   520(16)  & 2                     &  47--75    &  6.04-13.9             & \\ \hline
XTE 1859$+$226        &            &  0.380(3)             & 7.4(11)    &                        & g  \\
\ti{V406 Vulpeculae}  & 570(27)    &                       &            &                        & \\ \hline
XN Muscae 91          &  K4 V      & 0.4326                & 2.86--3.16 & 0.56--0.90             & c,h \\
\ti{GS\,1124$-$683}   &  406(7)    & 3.0                   & 54.0(15)   & 6.95(6)                & \\ \hline
XN Ophiuchi 77        &  K3 V      & 0.5213                & 4.44--4.86 &  0.3--0.6              & c\\
\ti{H\,1705$-$250}    &   420(30)  &  5.5                  &   60--80   &  5.2--8.6              & \\ \hline
IL Lupi               &  A2 V      & 1.1164                & 0.252(11)  &  1.3--2.6              & i,j \\
\ti{4U 1543$-$47}     &  129.6(18) &  9.1(11)              & $\sim$ 22  &  2.0--9.7              & \\ \hline
XTE J1550$-$564       & G8IV-K4III &  1.552(10)            & 6.86(71)   & 1.31$^{+0.33}_{-0.37}$ & k\\
\ti{V381 Normae}      & 349(12)    &  4.7-5.9 (?)          & 70.8--75.4 & 10.56$^{+1.02}_{-0.88}$& \\ \hline
XN Scorpii 94         &  F6III     & 2.6127(8)             & 2.64--2.82 & 1.1--1.8               & c,l \\
\ti{GRO\,J1655$-$40}  &  227(2)    & 3.2                   &   67--71   & 5.1--5.7               & \\ \hline
V4641 Sagittarii      &  B9III     &  2.817                & 2.74(12)   & 6.53$^{+1.6}_{-1.03}$  & m \\
\ti{XTE J1819$-$254}  &  211.0(31) & 9.59$^{+2.72}_{-2.19}$&            & 9.61$^{+2.08}_{-0.88}$ & \\ \hline
Cyg\,X-1              &  O9.7Iab   &  5.5996               & 0.25(1)    & $\sim$ 17.8            & n \\
\ti{1956$+$350}       & 74.7(10)   &  2.5                  &            & $\sim$ 10.1            & \\ \hline
V404\,Cygni           &  K0 IV     & 6.4714                & 6.02--6.12 & 0.57--0.92             & c,o,p \\
\ti{GS\,2023$+$338}   & 208.5(7)   & 2.2-3.7               &   52--60   & 10.3--14.2             & \\ \hline
GRS 1915$+$105        & K-MIII     &  33.5(15)             & 9.5(30)    &  1.2(2)                & q \\
\ti{V1487 Aquilae}    & 140(15)    &  12.1(8)              & 70(2)      &  14(4)                 & \\ \hline
\end{tabular}}
\end{center}
\label{tab1}
\end{table}

\begin{table}
\caption{Parameters for Nova Scorpii \protect\cite{Beer02}
 and V4641 Sgr \protect\cite{Orosz01}.}

 \vskip 3mm
 \begin{center}
{\begin{tabular}{lcc}
\hline
Parameter                    & Nova Scorpii    & V4641 Sgr \\
\hline
Orbital period (days)        &  2.623          & 2.817 \\
Black hole mass ($\msun$)    &  $5.4\pm 0.3$   & $9.61^{+2.08}_{-0.88}$ \\
Companion mass  ($\msun$)    &  $1.45\pm 0.35$ & $6.53^{+1.6}_{-1.03}$ \\
Total mass ($\msun$)         &        6.85     & $16.19^{+3.58}_{-1.94} $ \\
Mass ratio                   &        0.27     & $1.50\pm 0.13$ \\
Orbital separation ($\rsun$) &       15.2      & $21.33^{+1.25}_{-1.02}$   \\
Companion radius ($\rsun$)   &        4.15     & $7.47^{+0.53}_{-0.47}$ \\
Distance (kpc)               &        3.2      & $9.59^{+2.72}_{-2.19}$ \\
\hline
\end{tabular}}
 \end{center}
\label{tab2}
\end{table}

In the following sections, we shall argue that the correlation between
black-hole mass and period also has physical meaning: the shorter the
orbital period, the more rapidly rotating the helium star progenitor 
to the black hole. Rapid rotation centrifugally prevents some
fraction of the helium 
star to collapse into a black hole, resulting in a smaller black hole mass.
The reason why the correlation in Fig.~\ref{FIG1} is weak is that evolution
of the binary since formation of the black hole has washed out the 
relation. Properly, we should consider the correlation between pre-explosion
orbital period and post-explosion black hole mass. Much of our work
presented here is concerned with understanding the evolution of these
binaries, and using this knowledge to find the systems for which we can
reconstruct those parameters. Using that subset, we show much better
agreement between our model predictions and the observed
relation between reconstructed period and mass; this supports our
evolutionary model and has ramifications for the origin of GRBs.

\section{THE EVOLUTION OF SOFT X-RAY TRANSIENTS\label{evol}}

\subsection{Prior to the formation of the black hole\label{evol.gen}}

Following on the work of Brown et al.\cite{Brown96}, who
showed the importance of mass loss of helium stars in binaries
determines the final outcome of binary evolution,
Brown et al.\cite{Brown99C}, Wellstein \& Langer\cite{Wellstein99},
and Brown et al.\cite{Brown01A} showed that massive
helium stars could evolve into high-mass black holes only if they were
covered with hydrogen during most of their helium core burning era
(Case C mass transfer in binaries). In case A
or B mass transfer in binaries (Roche Lobe overflow in main-sequence
or red-giant stage) the Fe core that was left was too low in mass
to go into a high-mass black hole.
Brown et~al.\cite{Brown01A} showed that high-mass black holes
could be formed only if the mass was taken off of the black hole
progenitor after helium core burning was finished; i.e. Case C mass
transfer. 

Brown et al.\cite{Brown01C} showed that with the Schaller et al.
evolution, this could happen only in the neighborhood of ZAMS mass 
$20\msun$, definitely not at $25\msun$ and higher, and then only
in a narrow range of initial supergiant radii $\sim 1000-1175\msun$.
Because of the wind
losses, the mass transfer would begin as Roche lobe overflow only in Case B 
mass transfer, with these higher main sequence masses, because the giant
radii of these stars exceed their radii in the supergiant phase.
However, the helium star progenitors in at least three
of the evolved binaries seem to be too massive for the $20-23\msun$
ZAMS progenitors used by Brown et al.\cite{Brown01C}: the black
hole in V404 Cyg is probably at least 
$10\msun$~\cite{Bailyn98,Shahbaz94,Shahbaz96}, and the black hole in Nova
Scorpii is of mass $\sim 5.4\pm 0.3\msun$~\cite{Beer02}
and the mass loss in black hole formation is $\gsim 5\msun$~\cite{Nelemans99}
so that the progenitor of the helium star
must have been
$\sim 11\msun$.  {}From Table~\ref{tab2}, the black hole in V4641 Sgr is
of mass $9.61^{+2.08}_{-0.88}\msun$~\cite{Orosz01}.  The tentative
conclusion from the above is that at least these binaries with evolved
companions seem to have come from helium cores of $\sim 11\msun$, or ZAMS
mass $\sim 30\msun$.
With high wind mass loss rates as proposed by Schaller et~al.\cite{Schaller92},
such massive stars have larger radii as giants than as supergiants,
thus making Case C mass transfer impossible.  However, since radii
and mass loss rates of evolved stars are very uncertain, 
Lee et al.\cite{Lee2002} took the
view that the need for $\sim 11\msun$ helium cores implies that their
progenitors, $30\msun$ main-sequence stars, do expand enough to allow
Case C mass transfer. This was done by reduction of mass loss during He
burning, the extension being made by hand. 

Podsiadlowski et al.\cite{Pods02} studied mass deposition on the Nova Scorpii
companion very thoroughly. However, they used as black hole progenitor
a $16\msun$ He star, corresponding to a ZAMS $40\msun$ star. After
having modified wind losses by hand in order to extend progenitors
up to ZAMS $\sim 30\msun$, we are hardly in a position to criticize
this procedure, but we feel uncomfortable with high-mass black holes
following from such high ZAMS masses. Brown et al.\cite{Brown96}
calculated that 4U 1700-37, a low-mass compact object, had as
progenitor a ZAMS $40\pm 10\msun$ star.
Ankay et al.\cite{Ankay01} argue that the progenitor of the compact object
is $\gsim 35\msun$.
Reynolds et al.\cite{Reynolds99} modeled the X-ray spectrum of 4U 1700-37,
obtained with Beppo SAX and report the presence of a possible cyclotron
feature. If real, this observation yields a magnetic field of about
$5\times 10^{12}$G, so that 4U 1700-37 must be a neutron star.
Brown et al.\cite{Brown96} interpreted the low-mass compact object
from the region of ZAMS mass $\sim 40\msun$ as meaning that the envelopes
of stars in this mass range had such large wind loss rates that only
low-mass compact objects could be formed. In any case, Lee et al.\cite{Lee2002}
showed that Nova Scorpii could be evolved from a ZAMS $\sim 30\msun$
star, as long as wind losses were cut down sufficiently to enable
Case C mass transfer, so we suggest that stars as massive as ZAMS $40\msun$
may end up as low-mass compact objects. 

We note that a low-mass compact object would fit into the 
Brandt et al.\cite{Brandt95}
scheme of a two step process in which a neutron star with a kick velocity
is first formed, and then accretes enough matter to go into a black hole.
Podsiadlowski et al.\cite{Pods02} evolve Nova Scorpii from a different mechanism
than the other black hole SXTs, which have small system velocities,
as compared with the $106$~km~s$^{-1}$ of Nova Scorpii. We prefer to evolve
Nova Scorpii similarly to the other SXTs, although it is special because of its
large system velocity, which we interpret as indicating that it lost a
lot of mass in the explosion, substantially more than the other black
hole transient source binaries.

Some uncertainty in the evolution of all compact X-ray binaries is
the phase of spiral-in that occurred in their evolution: these binaries
are initially very wide, and when the primary fills its Roche lobe and
transfers mass to the secondary, the mass transfer leads to instability,
resulting in the secondary plunging into the primary's envelope. Next,
dissipation of orbital energy of the secondary causes the primary's envelope
to be ejected, and the orbit to shrink. Following the original work by
Webbink\cite{Webbink84}, Brown et al.\cite{Brown01C}
write the standard formula for common envelope evolution as
    \be
    \frac{G M_pM_e}{\lambda R} 
    = \frac{G M_pM_e}{\lambda r_L a_i} 
    = \alpha_{ce} 
     \left(\frac{G M_{\He} M_d}{2 a_f}-\frac{G M_p M_d}{2a_i}\right)
    \label{eq4}
    \ee
where $M_p$ is the total mass of the BH progenitor star just before
the common envelope forms, $M_e$ is the mass of its hydrogen envelope,
$M_{\He}$ is the mass of its core, $a_{\rm i}$ and $a_{\rm f}$ are the
initial and final separation, before and after the common envelope,
respectively, and $r_L\equiv R_L/a$ 
is the dimensionless Roche-lobe radius. This
equation essentially relates the loss of orbital energy of the secondary
to the binding energy of the ejected envelope. The parameter $\lambda$
is a shape parameter for the density profile of the envelope. It can vary
greatly between stars \cite{Tauris01}, but for the extended, deeply
convective giants we deal with in Case C mass transfer it is always
close to 7/6. (See also Appendix C of Brown et al.\cite{Brown01B}.)  
The parameter
$\alpha_{ce}$ accounts for the efficiency with which orbital energy is
used to expel the envelope, and may also account for some other effects
such as extra energy sources and the possibility that each mass element
of the envelope receives more than the minimum energy needed to escape
(see, e.g., Bhattacharya \& Van den Heuvel\cite{Bhatta91B} 
and references therein).

Given the parameters of the system
at first Roche contact, when spiral-in starts, the final separation is
determined by the product of $\lambda$ and $\alpha_{ce}$, 
the efficiency of the energy conversion. In general, these parameters are
only the simplest recipe prescription for the complex hydrodynamical 
interaction during spiral-in. While we therefore cannot predict the value
of $\lambda\alpha_{ce}$ from first principles, we can try to find its
value from constraints in some systems, and then assume it is the same for all
similar systems.

To calibrate the spiral-in efficiency, we need to
find systems in which we can also estimate the orbital separation
just after spiral-in well. This is complicated by the fact that mass
transfer has taken place since the spiral-in. 
Most SXTs have small
mass ratios, and for such small mass ratios the orbital separation
is fairly sensitive to the amount of mass transferred, making it
hard to derive the post-spiral-in separation from the present one.
The exception is V4641\,Sgr, in which the present mass ratio is close
to 1 (Lee et al.\cite{Lee2002}). 
Since the initial mass ratio could not have been significantly
greater than 1 (since that would result in unstable mass transfer), and
furthermore the orbital period changes very little with mass transfer
for nearly equal masses, we can fairly approximate the post-spiral-in
separation by the present one. Lee et al.\cite{Lee2002} showed the 
predicted ranges of post-spiral-in orbital periods for different values
of $\lambda\alpha_{ce}$, and a value quite close to 0.2 was
indicated. 
For 4U\,1543$-$47 (IL Lupi), they
found that it is near the boundary
between evolved and main-sequence evolution. To place it there,
they found from the reconstructed orbital
period in their Fig~6 
that $\lambda\alpha_{ce}\sim 0.2$ is also
consistent with the properties of this system.

Now if we look at Eq.~(\ref{eq4}), we see that $a_f$ scales almost
linearly with the donor (companion) mass $M_d$. The envelope mass $M_e$
is roughly $0.7 M_{giant}$ (Bethe \& Brown\cite{Bethe98}) and we use
    \be
    M_{He}=0.08 (M_{giant}/\msun)^{1.45} \msun
    \ee
so that
    \be
    a_f \propto \frac{M_d}{\msun} \left(\frac{M_{giant}}{\msun}\right)^{-0.55}
    a_i
   \label{eqaf}
   \ee
assuming $\lambda\alpha_{ce}$ to be constant and with neglect of the small
term in $a_i^{-1}$ in the r.h.s of Eq.~(\ref{eq4}).  {}From 
the possible ranges for the Case C mass transfer,
we note that the 20\% possible variation in $a_i$ results
in the same percentage variation in $a_f$. Because the
actual ZAMS mass can be anywhere in the range
$20-30\msun$ there can be an additional $\sim 25\%$ variation in $a_f$
with giant mass, as compared with the linear dependence on $M_d$. In
view of the modest size of these variations at a given donor mass, we make the
approximation in the rest of the paper that the pre-explosion orbital
separation depends only on $M_d$, and scales linearly with $M_d$.
This simple scaling and the modest amount of scatter around it are
partly the result of the weak dependences on initial parameters in 
Eq.~(\ref{eqaf}), but chiefly the result of the fact that our
model uses Case C mass transfer. This constrains the Roche contact to
first occur when the orbital separation of the binary 
is in a very narrow range, e.g., $\sim 1700\rsun$ for ZAMS $20\msun$ star
and $1\msun$ companion, as we make clear in the Appendix.

In short, the general properties of SXTs and the specific cases of
V4641\,Sgr and IL Lupi favor $\lambda\alpha_{ce}\sim 0.2$, which we 
therefore adopt as a general efficiency 
for the evolution of other transient sources. This then makes it possible
to make quite specific predictions for the prior evolution of many of the
other SXTs.

   \subsection{Expected regularities}

{}From the above theory, certain regularities follow for the system
behavior as a function of companion mass. First, the binding energy
relation for spiral-in (Eq.~(\ref{eq4})) shows that very nearly $a_f\propto
M_d$, with not much variation due to other aspects of the systems
(see previous section). Furthermore, the relation between Roche lobe
radius and donor mass when $M_d\ll M_{BH}$ implies that $R_L/a_f\propto
M_d^{1/3}$ (e.g., Eggleton\cite{Eggleton83}). As a result, the Roche lobe radius of
the donor just after spiral-in will scale with donor mass as $R_L\propto
M_d^{4/3}$. On the other hand, the donor radius itself depends on
its mass only as $R_d\propto M_d^{0.8}$.  Therefore, a low-mass donor
overfills its Roche lobe immediately after spiral-in. In the donor mass
range we consider ($M_d\gsim0.7\msun$) it does not overfill its Roche lobe
by much, so we assume that the system adjusts itself quickly by transfer
of a small amount of mass to the He star, which widens the orbit until
the donor fills its Roche lobe exactly.
Above this minimum mass, there will be a range
of donor masses that are close enough to filling their Roche lobes
after spiral-in that they will be tidally locked and will come into
contact via angular momentum loss (AML). 
Above this, there will be a range of mixed evolution,
where both AML and nuclear evolution (Nu) play a role. Finally, for the most
massive donors, $M_d>2.5\msun$, the post-spiral-in orbits will be too
wide for AML to shrink them much, so mass transfer will be initiated
only via nuclear expansion of the donor.  Of course, the ranges of Case
C radii of stars and variations of primary masses will ensure that the
boundaries between these regions are not sharp: near the boundaries the
fate of the system depends on its precise initial parameters.

\subsection{Comments on Case C mass transfer}
\label{sec2.1}

Brown et al.\cite{Brown99C} and Wellstein \& Langer \cite{Wellstein99} motivated
the need for Case C mass transfer, mass transfer following core
He burning, in order to ``trick" the companion in a binary into
evolving through the important part of its lifetime as a single star,
and then only removing its H envelope when that evolution was finished.
The point of the authors mentioned was that ``naked" He cores, evolved
in binaries in either Case A or Case B mass transfer, blew away,
ending up in Fe cores that were too low in mass to evolve into high-mass
black holes. 
We go through this argument in detail in the Appendix.


An extensive comparison of ``naked" and ``clothed" He star
evolution was carried out by Brown, et al.\cite{Brown01A}. 
The important question
of mass loss during the Wolf Rayet stage is addressed in this section.
Earlier calculations used mass losses deduced from Wolf-Rayet winds,
driven chiefly by free-free scattering. These depended on the square of
the matter density, and because of clumpiness were too large. As discussed
in Brown et al. Moffat \& Robert\cite{Moffat94} arrive at a wind loss rate
2.1 times smaller than that used by Woosley et al.\cite{Woosley95},
obtained from the polarization of the Thomson scattering which is linear
in the density. This value of wind loss is roughly consistent with that
measured from the period change in V404 Cyg.

The conclusion of Brown et al.\cite{Brown01A} was that even with the reduced
wind losses, enough of the naked He star would blow away that the
Fe cores following Case A or Case B mass transfer were insufficient
in mass to evolve into high-mass black holes.

Nelemans \& van den Heuvel\cite{Nelemans01} use mass loss rates by Nugis
\& Lamers\cite{Nugis00} which are substantially lower than those used by
Wellstein \& Langer\cite{Wellstein99} in the evolution of the CO cores evolved
by Heger and described in Brown et al.\cite{Brown01A}. They obtain much more
massive final He star masses.

\begin{table}
\caption{Final He and Fe core masses evolved from a ZAMS $60\msun$ star.
Units are $\msun$'s.  $\star$~Mass loss reduction relative to 
Braun\protect\cite{Braun97}.}
\vskip 3mm
\begin{center}
{
\begin{tabular}{cccc}
\hline
Mass Loss Rate$^\star$ &  Final $M_{He}$ & Final Fe core & Remnant mass \\
\hline
  1/1     &  3.132 & 1.352 & 1.35 \\
  1/2     &  4.389 & 1.305 & 1.17 \\
  1/3     &  6.108 & 1.606 & 2.11 \\
  1/4     &  7.550 & 1.749 & 5.20 \\
  1/6     &  10.75 & 1.497 & 10.7 \\
\hline
\end{tabular}
}
\end{center}
\label{tabML}
\end{table}


However, the final Fe core mass is not a linear function of the final
He star masses, as discussed by Fryer, et al.\cite{Fryer01}. In Table~\ref{tabML}
we give results from Fryer    et al. for a ZAMS $60\msun$ star.
The different mass loss rates are scaled to those of Braun\cite{Braun97}. 
The Braun rate is essentially that of Woosley et al. \cite{Woosley95}. Fryer
et al. discuss the nonmonotonic nature of the Fe core masses as function of
He core in terms of the way in which C shell burning takes place.
Remnant masses include fallback following supernova explosions of
energies $\sim 10^{51}$ ergs. In our case of hypernova explosions of
much higher energies, provided by magnetohydrodynamical effects, we
expect little fallback.
However, the results of Fryer et al. do support the suggestion of Nelemans
\& van den Heuvel \cite{Nelemans01} that if mass loss rates are decreased sufficiently,
high mass black holes can be formed in supernova explosions following
Case A or B mass transfer, because of fallback. We believe that Case C mass
transfer is necessary in the evolution of high mass black holes in the
transient sources, however. Indeed, from the dynamics of the situation
resulting from the low mass companion, De Kool et al.\cite{DeKool87} evolved
A0620 in Case C mass transfer. Regularities that result from Case C mass
transfer seem to be realized in the transient sources with main sequence
companions\cite{Brown01C}. In hypernova explosions the
magnetohydrodynamics effects should be sufficient to expel matter that does
not immediately fall into the black hole, so that fallback will be rather
different from that in supernova explosions as outlined by Lee et al.\cite{Lee2002}.

It is clear that with low metallicity winds can be near the lower end of
those in Table~\ref{tabML}. We believe that it is no coincidence
that there are two continuously shining black hole binaries LMC X-1 and
LMC X-3 in the LMC which has low metallicity, 
whereas only Cygnus X-1 is known in the Galaxy.

\section{RECONSTRUCTING THE PRE-EXPLOSION ORBITS\label{preexp}}
\label{reconst}

\subsection{Evolution of SXTs with evolved companions}

\underline{\it Nova Sco 94 (GRO\,J1655$-$40):\/}

The most extensive evolutionary studies have been made for Nova Scorpii.
Starting from the work of Reg\H{o}s et al.\cite{Regos98} who
make the case that the companion is in late main sequence evolution,
Beer \& Podsiadlowski\cite{Beer02} carry out extensive numerical calculations of the
evolution, starting with a pre-explosion mass of $2.5\msun$ and separation
of $\sim 6\rsun$. More schematically we arrived at a pre-explosion
mass of $1.91\msun$ and separation of $5.33\rsun$ \cite{Lee2002}.
We consequently have an 0.4 day pre-explosion period. With $\sim 6\msun$
mass loss in the explosion\cite{Nelemans99}, nearly half the system
mass, the binary period increases to 1.5 day, well beyond the period gap. 
This is also the period required if the common-envelope efficiency
in this binary was again 0.2 \cite{Lee2002}.
This explains why Nova Sco is the only system with a 
black-hole mass in the lower end of the range: its evolution really places
it among the narrow-orbit systems. Generally, the mass loss during explosion
is mild, and does not change which category a system belongs to. But in
those exceptional cases where the mass loss comes close to half the
total mass, the orbit widens very much and converts an AML system to a
nuclear-evolution system.
After explosion the binary evolves to its 
present period by nearly conservative mass transfer. Our estimate is
that $0.41\msun$ is transferred from the donor to the black hole.
Brown et~al.\cite{Brown99D} first made the case that Nova Scorpii was the
relic of a GRB.

\noindent
\underline{\it IL Lupi:\/}

IL Lupi may be one of
the most interesting binary after Nova Scorpii; it is quite
similar to the latter. Its present period is just at middle of the
period ``gap", i.e., binaries with the same companion mass but shorter
periods will lose angular momentum by gravitational waves and magnetic
braking, shortening their periods as they lose angular momentum, whereas
those with longer periods will move outwards, widening their orbit,
as they evolve transferring mass to the black hole. Orosz et al.\cite{Orosz98}
find the companion to be a late (estimated $3.93\times 10^8$ yrs post ZAMS)
main sequence A-star.
Recently over-abundances of Mg in the companion star of IL Lupi have been
observed \cite{Orosz02A}. In analogy with the case of the overabundances
in Nova Scorpii\cite{Israelian99,Lee2002,Brown00},
this indicates that there was an explosion at the time of black hole
formation in this system, in which some of the material ejected from
the core of the helium-star progenitor to the black hole ended up on
the companion.  Based on these observations and our given efficiency
$\lambda\alpha_{ce}=0.2$, one can start with $11\msun$ He star and
$1.7\msun$ companion as a possible progenitor of IL Lupi.  {}From the lower
boundary of the curve with $\lambda\alpha_{ce}=0.2$ in  Fig.~6
of Lee et al.\cite{Lee2002}
the period would be 0.5 days.  By losing $4.2\msun$ during the explosion,
the binary orbit would be widened to 1.12 day.  The period had to be
shortened to 0.8 day by magnetic braking and gravitation wave radiation
before the mass transfer started.  Conservative transfer of $0.23\msun$
from the companion to the black hole would bring the period from 0.8
day to the present 1.1164 day.

\noindent
\underline{\it V4641\,Sgr:\/}

As we discussed in section~\ref{evol}, this system is our calibrator 
for the spiral-in efficiency $\lambda\alpha_{ce}=0.2$ 
(Lee et al.\cite{Lee2002}).
The present mass ratio of V4641 Sgr is close to 1. 
Since the initial mass ratio could not have been significantly
greater than 1 (since that would result in unstable mass transfer), and
furthermore the orbital period changes very little with mass transfer
for nearly equal masses, 
we assume that its present state is
very close to the one immediately following spiral-in \cite{Lee2002}.

\noindent
\underline{\it GRS 1915$+$105:\/}

Recently Greiner et al.\cite{Greiner01} have determined the period and black hole
mass of GRS 1915$+$105 to be 33.5 day and 14$\pm 4\msun$. Interestingly,
we can evolve a system with properties very close to this by simply starting
from V4641\,Sgr and following its future evolution with conservative
mass transfer ($P_{orb}\propto\mu^3$, where $\mu$ is the reduced mass);
allowing for $4.6\msun$ to
be transferred from the donor to the black hole, we have
   \be
   P_{1915}=\left(\frac{9.61\times 6.53}{14.21\times 1.93}\right)^3
   P_{4641} = 33.7 \; {\rm day} .
   \ee
This would give a companion mass of $1.93\msun$, as compared with the
Greiner et al.\cite{Greiner01} mass of $M_d=1.2\pm 0.2\msun$.
However, the mass transfer cannot be completely conservative because of loss
by jets, etc., as evidenced by the microquasar character of this object. 
Furthermore the above $M_d$ is viewed as a lower limit by Greiner
et al. because the donor is being cooled by rapid mass loss, but its mass is
estimated by comparison with non-interacting stars. We thus believe our
evolution to be reasonable. We position the pre-explosion period and black hole
mass of GRS 1915$+$105 at the same point as V4641 Sgr. Since mass transfer
and widening of the orbit always occur together, the effect of this 
post-explosion evolution is to
introduce a weak secondary correlation between orbital period and companion
mass in the long-period regime, where such a correlation is not expected to
arise from the pre-explosion evolution.

\noindent
\underline{\it V404\,Cyg:\/}

The black hole in V404 Cyg appears to be somewhat more massive than in IL Lupi,
so we begin with a similar mass companion, but a $10\msun$ black hole, which
would have a period of 0.63 day. Again, we neglect mass loss in the explosion,
although a small correction for this might be made later. Conservative
transfer of $1\msun$ from the donor to the
black hole then brings the period to
   \be
   0.63 \; {\rm day} \left(\frac{1.7\msun\times 10\msun}{0.7\msun\times 11\msun}\right)^3 
   = 6.7\;  {\rm day}
  \ee
close to the present $6.47$ day period. Here we take $11\msun$ and $0.7\msun$
as current masses in V404 Cyg \cite{Orosz02A}.
The black hole in V404 Cyg seems to be somewhat more massive than the others
in the transient sources, with the exception of that in GRS 1915$+$105. 
In both cases we achieve the relatively high black hole masses
and periods by substantial accretion onto the black hole.

\noindent
\underline{\it GRO J1550$-$564:\/}

\begin{figure}
\centerline{\epsfig{file=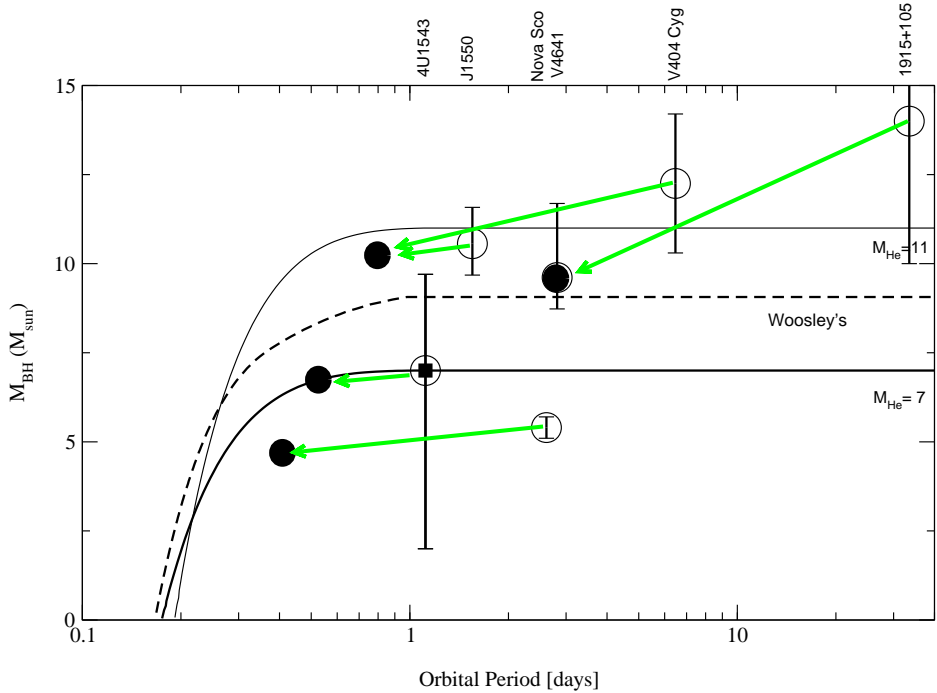,height=4in}}
\caption{Reconstructed pre-explosion orbital period vs.\ black hole masses 
of SXTs with evolved companions.
The reconstructed pre-explosion orbital periods and
black hole  masses are marked by filled circles, and
the current locations of binaries with evolved companions are marked by 
open circles.  
}
\label{FIG11}
\end{figure}

The high mass black hole in J1550$-$564, $10.56\msun$~\cite{Orosz02B},
is slightly 
less massive than the assumed black hole mass of V404 Cyg, 
and the companion is more massive than V404 Cyg with short
period, 1.552 days. So, we start from the same initial
conditions just derived for V404\,Cyg (Fig.~\ref{FIG11}), and end up
with the present system via simple conservative mass transfer.

\noindent
\underline{\it Cygnus X-1:\/}

Cyg\,X-1 is usually not considered to have come from the same evolutionary
path as the SXTs, since it is a persistent X-ray source with a much more
massive donor. But with the discovery of objects with relatively massive
donors in the SXT category, such as V4641\,Sgr, it is worth considering the
implications of our model for it.
Cyg X-1\,has been shown to have an appreciable system velocity\cite{Kaper99}
although it may be
only $1/3$ the $50$ km s$^{-1}$ given there, depending on the O-star
association (L. Kaper, private communication). The evolution of Cyg X-1
may have been similar to that of the transient sources, the difference
being in the copious mass loss from the companion O9I star, causing the
black hole to accrete and emit X rays
continuously. If we scale to Nova Scorpii to obtain the
initial binary separation, we find
   \be
   a_f =\frac{17.8\msun}{1.91\msun}\times 5.33 \; \rsun =50\; \rsun
   \ee
somewhat larger than the present binary separation of $40\rsun$. (We would
obtain $38\rsun$ if we scaled from the Beer \& Podsiadlowski \cite{Beer02} companion mass
of $2.5\msun$ for Nova Scorpii.)
Given uncertainties in the mass measurements, we believe it possible for
Cyg X-1 to be accommodated in this scheme. Some sort of common envelope
envelope evolution seems to be necessary to narrow the orbit in the
evolution involving the necessarily
very massive progenitor stars\cite{Brown01A}.

\subsection{Discussion of the hypercritical accretion onto black holes}

If we assume the time scale for the transfer,
$\sim 4\times 10^5$ yr for a $6.5\msun$ giant\cite{Schaller92,Belczynski02},
we would need a transfer
rate of $1.2\times 10^{-5}$ yr$^{-1}$, roughly 240 times Eddington limit
for an $10\msun$ black hole, where
the Eddington limit is $\dot M_{\rm Edd}=8\pi G M/\kappa_{es}c
\sim 5\times 10^{-8} \msun\; {\rm yr}^{-1}$ and
$\kappa_{es}$ is opacity. 

It seems that the rate is extremely high.
We would make the point, however, that hypercritical inflow,
in which the photons are carried in with the adiabatic inflow,
is much easier in black holes where the matter can easily
go over the event horizon, than onto neutron stars, where
accretion is onto the surface. The trapping radius for
hypercritical accretion was derived in Eq.~(2.10) of Brown\cite{Brown95}
\be
   r_{\rm tr} \simeq 0.6 R_s\dot m
\ee
where $R_s$ is the Schwarzschild radius and $\dot m=\dot M/\dot M_{\rm Edd}$.
Although this looks quite small,
for accretion onto the neutron star $r_{\rm tr}$ must be taken
to be much larger. That is, as described in 
Brown et al.\cite{Brown00}
in some detail,
the accreting matter is initially of too low temperature to cool
by neutrinos, and it simply accumulates on the neutron star, forming
an accretion shock.
The accretion shock has to pile up a lot of $\gamma=4/3$ matter
so that the central region gets hot enough to emit neutrino pairs.
Now the incoming matter can only be trapped outside of the accretion
shock so that $r_{\rm tr}\gsim r_{\rm sh}$, where\cite{Brown95}
  \be
  r_{\rm sh}\simeq 2.6\times 10^8 \; {\rm cm}\;
  (\dot M/\msun \; {\rm yr}^{-1})^{-0.37}
  \ee
It is this shock radius, which is relatively large compared with
the dimensions of the neutron star that make the minimum rate for
hypercritical accretion as large as
  \be
  \dot m\simeq 10^4  \;\;\;\;\; {\rm for\; neutron\; star}.
  \ee
In the case of the black hole accretor the matter can flow adiabatically
across the event horizon, once its angular momentum has been brought
down as described above by the accretion disk.

Hypercritical accretion in an accretion disk is somewhat different
from the spherical situation and we may have to be more careful
than in Brown et al.\cite{Brown00} where the accretion rate of
$\sim 1\; \msun$ yr$^{-1}$ meant $\dot M\sim 10^8\; \dot M_{\rm Edd}$
and the photons were clearly trapped and carried in with the adiabatic
inflow. Here we consider 100$-$1000 $\dot M_{\rm Edd}$.
With an accretion disk present, the photons will be trapped when the
time to carry them into the center of the accretion disk is less than
the time for the photons to diffuse out nearly spherically, since
the disk will be both geometrically and optically thick,
with nearly equal $H$ and $R$.

In Brown et al.\cite{Brown00}
we found that the inward radial velocity in the disk was
  \be
   |v_r|=(3/7) \alpha v_K
  \ee
where $v_K$ was the Keplerian velocity $\sqrt{GM/r}$,
a factor of $\sqrt{2}$ less than the free fall velocity
assumed for the photons. 
The viscous disk time is, thus,
  \be
  \tau_{\rm visc}=\frac{r}{|v_r|}.
  \ee
For the electron diffusion the optical depth is
  \be
  \tau_{es}=\int^\infty_r \rho \kappa_{es} dr
  =\frac{2 R\dot m}{\sqrt{r r_s}}
  \ee
where $ r_s = 2GM/c^2$ and $\dot M_{\rm Edd}=4\pi c R/\kappa_{es}$.
This gives, in random walk, a dynamical time for diffusion of
  \be
  \tau_{\rm diff} \simeq \frac{h}{c}\frac{h}{\lambda_{es}}
  =\frac{h^2}{c}\kappa_{es}\rho.
  \ee
We obtain the trapping radius by setting
  \be
  \tau_{\rm visc} =\tau_{\rm diff} 
  \ee
where we have used the random walk diffusion time, we
find the trapping radius
  \be
  r_{tr}= 0.5 \dot m R,
  \ee
in agreement with Begelman \& Meier\cite{begelman82}.
This should be compared with $r_{tr}=2 R\dot m$ for the
spherical situation\cite{Brown94B}, which was lowered to
$r_{tr}=0.6 R_s\dot m$ in Brown\cite{Brown95}. The decrease came
from replacing random walk by a solution of the diffusion equation.

We\cite{Bethe98,Brown95} pointed out that in common envelope evolution
a neutron star can accrete $\sim 1\msun$ yr$^{-1}$, because in
hypercritical accretion the photons are swept in by the adiabatic
inflow. 
However, for hypercritical accretion a lower limit
of $\dot M=10^4\dot M_{\rm Edd}$ is needed. This results from the
pressure of the radiation, as does the usual limit of $\dot M_{\rm Edd}$.
However, if there is no surface onto which the material accretes
and radiates away its binding energy but just the event horizon
as on a black hole, this lower limit need not be applicable. The
matter can simply be swept over the event horizon in the adiabatic
inflow.

The matter must first be ``processed", i.e., have most of its angular
momentum removed, before it can go into the black hole. This ``processing"
should be carried out by the accretion disk. One might think that
the luminosity from the disk should not exceed Eddington; otherwise
the pressure might remove the outer parts of the disk.
In Grimm et al.\cite{Grimm01} the average luminosity for 1915$+$105
is that of a hydrogen accretion rate of $\sim 3\times 10^{-8}\msun$
yr$^{-1}$ onto a neutron star, whereas that for V4641 Sgr
is $\sim 100$ times less. In both cases there are jets, indicating
super Eddington accretion at times.

We interpret the jets to result from the accretion onto the
accretion disk being so great at times that the accretion disk
cannot accept it all. That 1915$+$105 is much more active than
V4641 Sgr is quantified by the $\sim 100$ times greater average
luminosity. However, at high accretion rates we would expect the
disk to be optically thick and that the photons would be carried
by the matter over the event horizon so there may be some limit
to the disk luminosity, with increasing accretion.

As Greiner et al. wrote \cite{Greiner01}, the black hole in 1915 may be
spinning rapidly. We\cite{grb2000} show that generally
$\sim 10^{53}$ ergs is available in the spin energy of the
black hole, but that only $\sim 10^{52}$ ergs is used in the
GRB and hypernova explosion, mostly in the latter.

We believe that in the short time $\sim 5-10$ sec central engines
enough energy will be transferred from the rapidly rotating black hole
to the accretion disk to power the hypernova explosion. The latter
dismantles the accretion disk so that no further rotational energy
can be transferred, leaning the black hole in rapid rotation.

\subsection{Problems with the close (AML) systems\label{preexp.aml}}

Reconstruction of the AML binaries is more complicated, because they have 
lost angular momentum through magnetic braking and gravitational waves,
so that their present positions as plotted in Fig.~\ref{FIG1} are not
those at pre-explosion time. As with the evolved companions, matter will
have been accreted onto the black hole, so the black hole masses will
be somewhat greater than just following the explosion.
As noted earlier, the binaries with less massive companions with
separation $a_f$ at the end of common envelope evolution overfill their
Roche Lobes. The outer part of the companion, down to the Roche Lobe
$R_L$ is transferred onto the He star. This mass transfer widens the
orbit to $R_L$, possibly overshooting. Unless much mass is lost in
the explosion when the black hole is formed, the Roche lobe radius
is unchanged by the formation of the black hole, and corresponds to
line III in Fig.~\ref{FIG2}.

\begin{figure}
\centerline{\epsfig{file=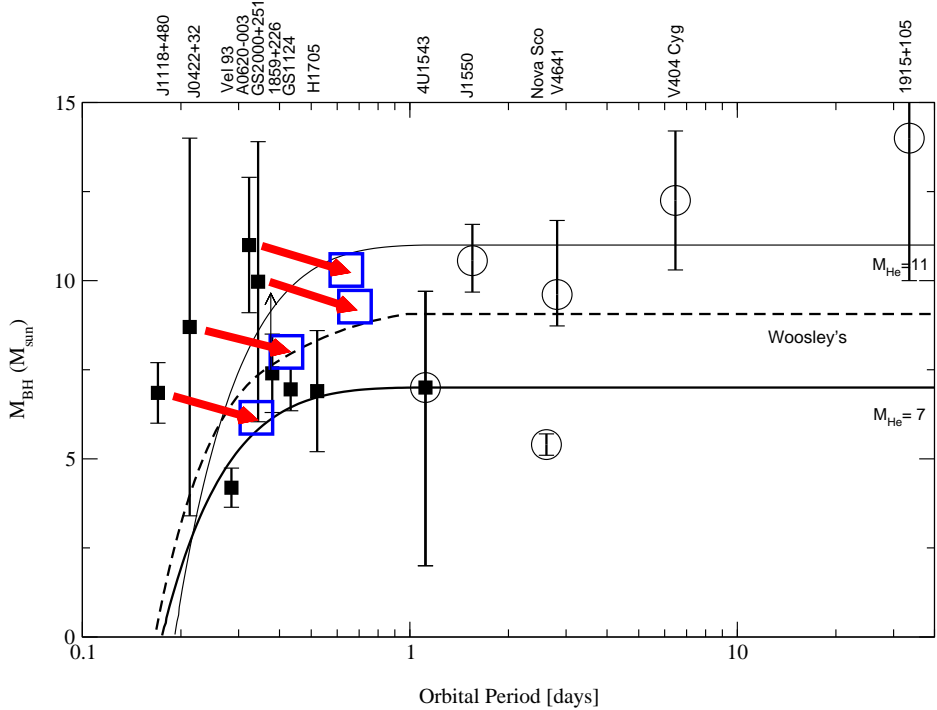,height=4in}}
\caption{Present orbital period vs.\ black hole masses 
of SXTs.  The deviations from the theoretical
curves are substantial due to the post-explosion
evolution of the binaries. The arrows on the AML systems point to an
approximate post-explosion location if the donor mass was initially 
1.5$\msun$, and 0.7$\msun$ has now been transferred to the black hole.
The solid lines 
indicate the possible ranges of black hole masses with 
polytropic index $n=3$ (radiative), for given
pre-explosion spin periods which are assumed the same as the pre-explosion
orbital period. Here we used $R_{He}=0.22 (M_{He}/\msun)^{0.6} \rsun$. 
For comparison, the results with a ``scaled" He core
($9.15\msun$) of Woosley's $25\msun$ star at the beginning of $^{12}$C
burning with $T_c= 5\times 10^8 K$, appropriate for Case C mass transfer, are
plotted as a dashed line\protect\cite{Woosley01}. 
In this plot, we scaled the radius of Woosley's
core, $R_{\rm Woosley}\sim 3\times 10^{10}$\,cm, 
by a factor 2. As can be seen, the
AML systems can plausible originate from systems within the curves, and
thus are consistent with our theory. However, since they could have originated
anywhere between the open square and their current location, they do not
strongly test the theory.
}
\label{FIG10}
\end{figure}

\begin{figure}
\centerline{\epsfig{file=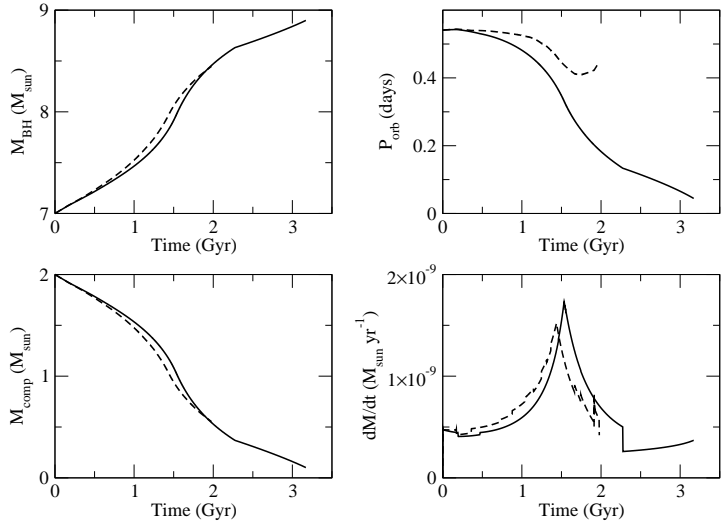,height=3.5in}}
\caption{Evolution of a binary with $7\msun$ black hole 
and $2\msun$ companion for the initial period of 0.54 day.
The solid line marks the evolution in case the companion star 
adjust itself as it loses mass;
the dashed line traces the evolution in case the mass loss does not
affect the internal time scale of the companion star, so that it
follows the same time evolution as an undisturbed $2\msun$ star.
}
\label{FIG7}
\end{figure}

\begin{figure}
\centerline{\epsfig{file=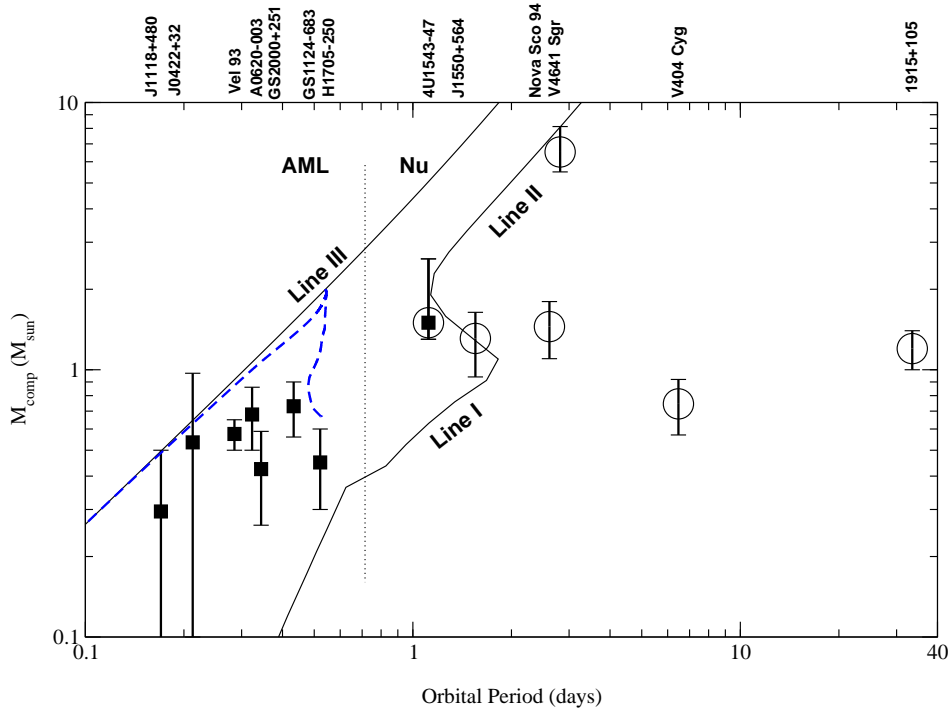,height=4in}}
\caption{Evolutionary tracks of a binary with $9\msun$ black hole 
and $2\msun$ companion for the initial period of 0.54 day.
The two evolution possibilities are as in Fig.~\ref{FIG7}.
Left (right) dashed line corresponds to the solid (dashed)
lines in Fig.~\ref{FIG7}.
}
\label{FIG8}
\end{figure}

Brown et al.\cite{Brown01C} explored the evolution of ZAMS
$1.25\msun$ stars under magnetic braking, gravitational waves and
mass transfer to the black hole. We adapt the same methods to make
a more detailed study of the AML.
First of all we construct (Fig.~\ref{FIG2}) the lower limit on the
companion mass for evolution in a Hubble time, giving the dot-dashed line
there. All binaries  with companions in main sequence at the beginning
of mass transfer must lie between the dot-dashed line and line III
in that figure.  The fact that the AMLs tend to lie below the dot-dashed
line implies both mass loss from the companion and accretion onto the
black hole. Therefore, all these systems have shrunk their orbits and
increased their black-hole mass since the formation of the black hole,
by amounts that cannot be determined well.
In Fig.~\ref{FIG10}, however, we show where the four shortest period
AMLs would have come from, had they lost 0.7~$\msun$ from an
initial 1.5~$\msun$.
{}From our earlier discussion about the $a_f$ following common envelope
evolution we saw that binaries with companions which stayed in main
sequence were favored to come from companion masses less than $2\msun$,
and from Fig.~\ref{FIG2}, we see that they would chiefly have
companion ZAMS mass greater than $\sim 1\msun$, so that most of them would
initially have periods of 0.4--0.7 days (which follows from the separations
obtained from our Eq.~(\ref{eqaf})).
In trying to understand the detailed evolution of the AML we begin
from a binary with a $2\msun$ companion which just fills its Roche Lobe
following common envelope evolution. We then follow its evolution under
the two different assumptions made in Brown et al.\cite{Brown01C}: (1)
That its time of evolution is always given by its initial $2\msun$ mass,
i.e., ignoring effects of mass loss on the internal evolution
time (dashed lines in Fig.~\ref{FIG7}
and right dashed line in Fig.~\ref{FIG8}) 
(2) That the evolution of the star proceeds according
to its adjusted mass (solid lines in Fig.~\ref{FIG7} and left
dashed line in Fig.~\ref{FIG8}).
Since mass loss drives the companion out of thermal equilibrium, these
two extremes bracket the outcome of a full stellar model calculation.

In summary, the AML systems have had the information on their
post-explosion parameters partly erased by subsequent evolution, in a 
manner that we cannot undo. Therefore, they can only provide a crude
consistency check on the mass-period relation for black holes in SXTs,
rather than provide precise constraints.

\section{ANGULAR MOMENTUM AND ITS CONSEQUENCES FOR THE MASS AND SPIN OF
         THE BLACK HOLE}
\label{rotate}

It is, in general, a difficult and unsolved problem to calculate the
angular momentum of a stellar core at any given time. Even if we make the
usual assumption that the rotation is initially solid-body, and not very
far away from the maximal stable rotation frequency, the viscous coupling
between the various layers of the star as it evolves is poorly known, and
thus it is hard to be very quantitative. The general trend, however, is that
the core will shrink and the envelope expand. In absence of viscous coupling,
every mass element retains its angular momentum, and hence the core spins
up as the envelope spins down, setting up a strong gradient in rotation
frequency between the core and the envelope. Viscosity will then act to reduce
this gradient, transporting angular momentum from the core to the envelope,
but the efficiency of this process is very uncertain\cite{Spruit98,Livio98}.

As we noted above (section 1), in our scenario, a number of effects 
will increase the angular momentum of
the core relative to a similar core of a single star: (1)
during spiral-in, the matter somewhat inside the orbit of the secondary is
spun up by tidal torques\cite{Rasio96}; (2) the removal of the
envelope halts the viscous slowdown of the core by friction with the envelope;
(3) during the post-spiral-in evolution, tidal coupling will tend to spin
the helium star up even closer to the orbital period than was achieved by
the first effect. This will not be a very strong effect because the duration
of this phase is short, but it will affect the outer parts of the
helium star somewhat, and this
is the most important part (see below).

The net result of all these effects will be that the helium star will spin
fairly rapidly, especially its envelope. The core is not so crucial to our
argument about the fraction of the star that can fall into the black hole,
since the few solar masses in it will not be centrifugally supported even
in quite short orbits. For the purpose of a definite calculation, we
therefore make the following assumptions: (1) the helium star co-rotates 
with the orbit before explosion and is in solid-body rotation; 
(2) the mass distribution of the helium
star with radius is given by a fully radiative zero-age helium main sequence
star. This latter approximation is, of course, not extremely good. However,
what counts is the angular momentum as a function of mass, so the fact that
the mass distribution has changed from helium ZAMS to explosion would be
entirely inconsequential if no redistribution of angular momentum had taken
place in the interim. As we saw above, any redistribution of angular momentum
would take the form of angular momentum transport toward the outer layers.
This means that relative to our ideal calculations below, a better calculation
would find more angular momentum in the outer layers, and therefore
somewhat smaller black hole masses than the ones we calculate.

\begin{figure}
\centerline{\epsfig{file=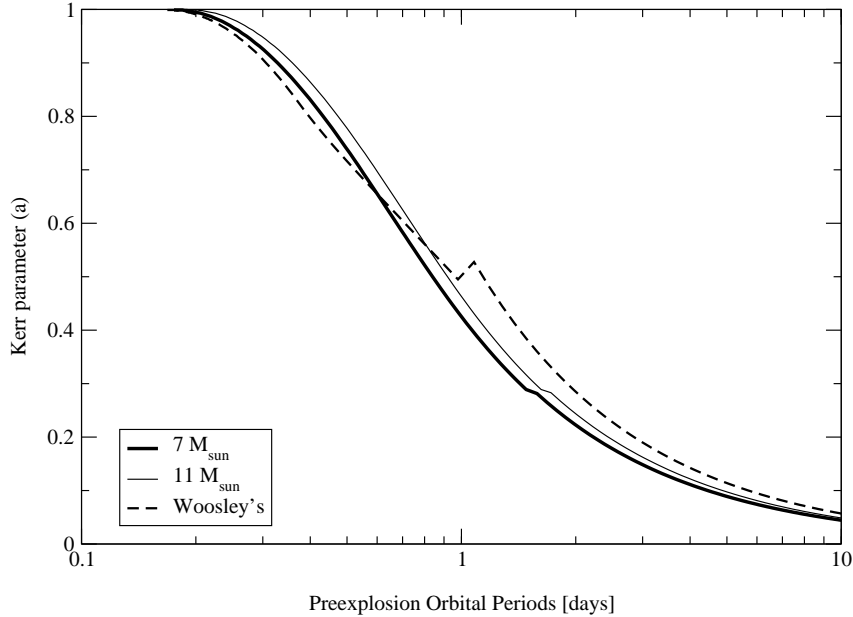,height=4in}}
\caption{The Kerr parameter of the black hole resulting from the collapse
of a helium star synchronous with the orbit, as a function of orbital period.
The conditions are the same as in Fig.~\ref{FIG10}, as is the meaning of the
three curves. 
Woosley's helium core is of mass $9.15\msun$ from a ZAMS $25\msun$ 
star\protect\cite{Woosley01}.
Note that the result depends very little on the mass of the
helium star, or on whether we use a simple polytrope or a more sophisticated
model. The plot illustrates that rapidly rotating black holes needed for
powering GRBs originate only from originally short-period SXTs.
\label{FIG12}
}
\end{figure}

We now investigate how much mass will be prevented from falling
into the black hole by the angular
momentum of the He star, under the above assumptions of solid-body rotation
with a period equal to that of the binary.
If we assume that angular momentum is conserved during the
collapse, we can get the cylindrical radius $R_c$ within which matter is
not centrifugally prevented from falling into the black hole:
   \be
   R_c^2\Omega =  \tilde l(\hat a) \frac{G M_c}{c} 
   \ee
where 
$\tilde l(\hat a)$ is the dimensionless
specific angular momentum of the marginally bound
orbit for a given Kerr parameter $\hat a$, and
$M_c$ is the total mass inside the cylinder of radius $R_c$.
The Kerr parameter becomes
  \be
 \hat a= \frac{I_c\Omega}{G M_c^2/c} = k^2 \tilde l(\hat a)
  \ee
where $I_c$ is the total moment of inertia inside the cylinder of radius $R_c$,
$I_c = k^2 M_c R_c^2$.
$M_c$ gives an estimate of the final black hole mass. Combining these
relations with a profile of angular momentum and mass versus radius 
using the assumptions listed above, we can calculate the expected black hole
mass and Kerr parameter as a function of SXT period before explosion.

In Fig.~\ref{FIG10} we show the predicted relation between orbital period
and black-hole mass for different helium star masses in our model. We
compare these with the {\it present\/} properties of all SXTs for which
the required parameters are known. The properties are consistent with 
the theoretical relations, but do not confirm it very strongly due to the
evolutionary changes discussed in section~\ref{preexp}. 
Specifically, the AML systems lie
above and to the left of the curves, because their orbits shrunk and their
black holes accreted mass since the formation of the black hole.
However, as we saw in section~\ref{preexp.aml}, plausible amounts of
conservative mass transfer since the explosion would place the systems
among the theoretical post-explosion curves (indicated by the open squares
and arrows).

To test the theory more strongly, we show in Fig.~\ref{FIG11} only those
systems for which the pre-explosion properties could be reconstructed
(section~\ref{preexp}).  We compare the observed
points with ideal polytropic helium stars of 7\,$M_\odot$ and 11\,$M_\odot$,
and with a full model calculation obtained from Woosley\cite{Woosley01}. 
By coincidence, the
curves converge near the region of the shortest-period observed systems, so that
the uncertainty in helium star mass is not of great importance to the outcome.
A helium star mass in the lower end of the range (7--9\,$M_\odot$) may be
somewhat preferred for these systems. For periods above 1 day, angular-momentum
support is not important, and the 
mass of the final black hole will be very close to that of the helium star,
and thus varies somewhat from system to system. As we can see, the 
reconstructed pre-explosion properties lie much closer to the theoretical
predictions.

As a corollary, we find that systems with very large velocities, like Nova 
Sco, will be rare: at the shortest pre-explosion orbits, where much mass
is ejected, the companion mass tends to be small. Then the center of mass
of the binary is close to that of the helium star, which strongly limits the
systemic velocity induced by the mass loss. On the other hand, for the
widest systems, where the companion tends to be massive enough to allow a
significant systemic velocity induced by mass loss, the mass loss itself
becomes too small to induce much of a systemic velocity.

An important result for our proposed relation between SXTs and hypernovae
and GRBs is shown in Fig.~\ref{FIG12}. This figure shows the 
expected Kerr parameter of the black hole formed in our model. We see that
for the short-period systems, this Kerr parameter is very large, 0.7--0.9.
This means that we are justified in adding only the mass that immediately
falls in to the black hole, because as soon as the rapidly rotating black hole
is formed, it will drive a very large energy flux in the manner described
by Brown et al.\cite{grb2000}. 
This both causes a GRB and expels the leftover stellar envelope.
The systems with longer orbital periods do not give rise to black holes
with large Kerr parameters, and thus are presumably not the sites of GRBs.


\section{OTHER OBSERVATIONAL ISSUES}
\label{other}

\subsection{Deposition of metals in the companions of SXTs}
\label{sec4}

We now model the deposition of the expelled matter onto the
companion star in a highly schematic way. 

With a preexplosion separation in Nova Scorpii of $5.33\rsun$, and ejection
of $6\msun$ of matter, with an F-star radius of $1.68\rsun$, in a spherically
symmetrical explosion the fraction of solid angle subtended by the companion
at the time of explosion was
  \be
  \frac{\Omega}{4\pi} =\frac{\pi (1.68\rsun)^2}{4\pi (5.33\rsun)^2}
  =0.025
  \ee
which with $6\msun$ matter deposited would give 
  \be
  \Delta M_{dep}\sim 0.15\msun
  \ee
Because of the smaller separation of $5\msun$ and the greater companion mass
at the time of explosion, this is $\sim 5$ times greater than that estimated
by Brown et al.\cite{grb2000}. Less than a factor of 2 would now be needed, 
following
their estimates, for sufficient metal deposition off the companion.
As motivated there, such a factor should come from the asphericity of the
explosion.

We shall estimate the ratios of matter deposited on IL Lupi and V4641 Sgr
as compared with Nova Scorpii, assuming the factor for asymmetry enhancement
to be the same in these cases. Since the thermohaline time scale is short
compared with the lifetime of the companion star, we shall divide the amount
of deposited matter by the volume of the star in order to estimate the
anomaly in metal deposition.

The ratio of anomaly in IL Lupi to Nova Scorpii is thus estimated to be
  \be
  R_{\rm\footnotesize IL Lupi/Nova Sco} &=&
  \left(\frac{4.2\msun}{6\msun}\right)
  \left(\frac{1.91\msun}{1.7\msun}\right) 
  \left(\frac{1.53\rsun}{1.68\rsun}\right)^2
  \left(\frac{5.33\rsun}{6.19\rsun}\right)^2 \nonumber\\
  &\simeq& 0.48
  \ee
where the anomaly is assumed to be proportional to
  \be
  {\rm anomaly} \propto \left(\frac{\Delta M_{\rm explosion}}{M_d}
  \right) \left(\frac{R_d}{a}\right)^2.
  \ee
The ratio of anomaly in V4641 Sgr to Nova Scorpii is estimated as
  \be
  R_{\rm\footnotesize V4641 Sgr/Nova Sco} &=&
  \left(\frac{1.6\msun}{6\msun}\right)
  \left(\frac{1.91\msun}{6.53\msun}\right) 
  \left(\frac{4.49\rsun}{1.68\rsun}\right)^2
  \left(\frac{5.33\rsun}{21.3\rsun}\right)^2 \nonumber\\
  &\simeq& 0.04
  \ee
Our estimates thus show that the anomaly should be detectable in IL Lupi,
the oxygen abundance being $\sim 5$ times normal, but only marginally
so in V4641 Sgr, $\sim 40\%$.

\subsection{Ultraluminous X-ray source}

An interesting byproduct of our black hole evolution in binaries
is their possible and promising use to explain ultraluminous
X-ray sources (ULXs) in external galaxies. These have been extensively
observed by the Chandra X-ray observatory\cite{Makishima00}.
We follow here the discussion of King et al.\cite{King01}.
{\it ``A key to understand their nature may be that they appear to occur
preferentially, although not exclusively, in regions of star formation."}
The initial explanation of these ULXs was that they came from
$\sim 100\msun$ black holes; the disadvantage of this explanation was that
such objects are not seen in our Galaxy and no one knows how to evolve
them.

King et al.\cite{King01} suggest that the ULX's originate from the same sort
of binaries we have evolved above, the black hole SXTs.
The high luminosity arises when they cross the
Herzsprung gap, as GRS 1915$+$105 does at this time. The time scale for
this is the thermal time scale
  \be
  \tau_{\rm th}=\frac{3\times 10^7}{(M/\msun)^2}\; {\rm yr},
  \label{thermal}
  \ee
i.e. $7\times 10^5$ yrs for the $6.5\msun$ companion in V4641 Sgr which
becomes something like GRS 1915$+$105 after crossing much of the
Herzsprung gap according to our scenario\cite{Lee2002}.

The initial problem with the SXTs as explanation was that luminosities
of $\sim 10$ times Eddington (Here we use Eddington to mean that limit
for a neutron star with $\sim 20\%$ efficiency, midrange for the
efficiencies of 6 to 42\% for nonrotating and rapidly rotating black hole,
respectively.) were required.
The initially proposed $\sim 100\msun$ black holes
were brought down to $\sim 10\msun$ ones by the realization that
accretion could proceed at $\dot M\sim 10\dot M_{\rm Edd}$ because of
instabilities in the accretion disk which make the disk ``porous"
\cite{Shaviv98,Shaviv00,Begelman02}.
Since the Eddington limit goes linearly with the Schwarzschild radius
$R_s$ and $R_s$ linearly with the mass, $M_{BH}$, a $10\msun$
star accreting at $10\dot E_{\rm Edd}$ would look like a
$100\msun$ black hole accreting at Eddington.

Taking the Schaller et al.\cite{Schaller92} models we have estimated the
mass loss in going from V4641 Sgr to GRS 1915$+$105 in assumed
conservative mass transfer as in Lee et al.\cite{Lee2002}. 
The assumed $5\msun$
companion goes out of equilibrium in the mass transfer, which should
be calculated in an evolutionary code. We hope to get an estimate
by simply assuming the mass going over the Roche Lobe to be
transferred to the black hole, and readjusting the Roche Lobe
according to the changed mass ratio. We find the mass
transfer to be at a rate of $\dot M\sim 10^{-5}\msun$ per year,
about 200 times the $\sim 5\times 10^{-8}\msun$ yr$^{-1}$
assumed for Eddington.

Belloni et al.\cite{Belloni97} find the emission from 1915$+$105 to
be rather complicated in detail. They interpret the $\sim 1$ sec
appearance and disappearance of emission from an optically thick
inner accretion disk as coming from fluctuations in the inner
radius of the disk from $\sim 20-80$ km, the lower radius during
bursts and the upper during quiescence.

We cannot make a detailed model, but support that the accretion
is hypercritical most of the time, as would be necessary to build
up the black hole mass of 1915$+$105 from that V4641 Sgr over
$4-5\times 10^5$ years.

King et al.\cite{King01} emphasize the importance of beaming in building 
up the high apparent luminosities which are observed. Our evolution
of the SXTs as fossil remnants of GRBs involve initially large effects
of beaming in order to produce the GRBs, and, as noted
earlier, the binaries should be left in considerable rotation.
In fact, jets are seen in 1915$+$105, as well as evidence of rotation.

Our evolution\cite{Lee2002} where the separation $a_f$ between
donor (companion) and black hole following common envelope
evolution is proportional to $m_d$ (Eq.~(5) of Lee et al.\cite{Lee2002}),
is helpful in understanding which binaries are good progenitors
of ULXs. The ``silent partners" of Brown et al.\cite{Brown99C} become SXTs
only after they evolve, because only then can their
outer matter cross the Roche Lobe.
Their companions are the more ``massive" ones and consequently
pour more matter into the black hole.

We can obtain a limit on the companion mass necessary for a ULX
from the evolution of Nova Scorpii by Beer \& Podsiadlowski\cite{Beer02}.
Their F-star companion was initially of $2.5\msun$, and began
evolving in main sequence, with resulting loss of mass to the
black hole.
The calculation of Beer \& Podsiadlowski is a fully evolutionary one.
None the less we find that we can reproduce their results by
conservative mass transfer, in which the period of the binary
goes as the cube of the reduced mass. This is not surprising,
in that the quiescent luminosity in Nova Scorpii is
$\sim 2.5\times 10^{32}$ ergs, a factor of $\sim 10^6$ less
than Eddington\cite{Garcia97}.
During bursts the luminosity is higher, $\sim 10^{38}$ ergs.
What we learn from this is that rapid mass transfer -- although
less than Eddington -- is going on silently between companion
star and black hole while the observable effects are small,
except during outbursts. We extend this scenario now to the
Herzsprung gap, where both mass transfer and outbursts
are much greater.
The calculated rate of Beer \& Podsiadlowski\cite{Beer02} for the
(future) crossing of the Herzsprung Gap is $\sim 4\times 10^{-8}
\msun {\rm yr}^{-1}$, above $\dot M_{\rm Edd}$ for
their  $5.3\msun$ black hole.
According to Lee et al.\cite{Lee2002}, 
the reason that Nova Scorpii (and IL Lupi) can begin transferring
mass in main sequence is that they are kicked into relatively
large orbits in the explosion which forms the black hole due
to large mass losses. (Companions in both binaries show
substantial metal abundances that would have come from the
explosion.)

In general we found most of the binaries with initial
companion mass $<2.5\msun$ stay in main sequence\cite{Lee2002}
during the active stage of SXTs. 
We interpret
Nova Scorpii and IL Lupi as having companions in between
the main sequence ones and those of the silent partners,
as the companion mass increases. Thus, we estimate $m_d\sim 3\msun$
as minimum mass for the ``silent partner" companion.

A $3\msun$ companion could be stripped down to its $0.4\msun$
He core in transferring matter after expansion to the Herzsprung
gap. Using the thermal time scale of $3.3\times 10^6$ yrs this means
maximum average rate of mass transfer of $0.9\times 10^{-6}\msun$~yr$^{-1}$
or $\sim 20\dot M_{\rm Edd}$ for a $10\msun$ black hole.
Thus an actual rate of $\dot M\sim 10\dot M_{\rm Edd}$ is reasonable
for such a star, and progenitor binaries with $3\msun$
(and more massive) companion should give ULXs.

The highly super Eddington (hypercritical) accretion with more
massive companions, as in V4641 Sgr and 1915$+$105 is unlikely
to produce higher luminosity, because the high density infalling
matter can very efficiently trap the photons and carry them
adiabatically inwards with the adiabatic inflow. We propose to
divide the accretion disk into the outer thin part with $\dot m\lsim 10$
and the inner thick part with $\dot m \gsim 10$. (We basically
extend $\dot M_{\rm Edd}$ to $10\dot M_{\rm Edd}$ to take into
account the effects of inhomogeneities, assuming a ten fold increase.)
Photons in the outer thin disk are emitted; these in the inner
thin disk are trapped and carried inward. The two regions are
effectively separated by (approximately) stationary matter.

Given the complexities of accretion disks, this may seem to be
an oversimplified approximation. Indeed, the complication induced
by inhomogeneities in the thin disk has forced an increase in
$\dot M_{\rm Edd}$ to an effective $10\dot M_{\rm Edd}$,
where the factor 10 is an estimate from Begelman\cite{Begelman02}.
The concept of photon trapping is, however, an exact translation
of neutrino trapping in supernova explosions,
which is discussed in Bethe et al.\cite{Bethe79}. 
There are, however, instabilities
which lead to an oscillatory in the accretion, which we mention
below. A major assumption is that the superposition of these on our
above simple model does not change the average accretion rate.

Whereas SXTs such as 1915$+$105 may well produce many of the
observed ULXs, this cannot be the whole story, since they are
also observed in elliptical galaxies. King et al.\cite{King01} also suggest
unstable accretion during a thermal time scale of a radiative star
onto a neutron star such as occurs in Her X-1, the companion being
more massive than the neutron star, as a source. Whereas the
companions may be sufficiently massive in spiral galaxies to
produce the ULXs, it is not clear that they can do so in elliptical
galaxies. We have not dealt with the relevant binaries in this paper.

\subsection{Population synthesis of SXTs and GRBs}

The 14 SXTs in Table~\ref{tab1}
separate nearly equally into those with unevolved main sequence
and those with evolved companions. This immediately tells us that
there are many unobserved black-hole binaries, since the lifetime
for the evolved companions is nearly two orders of magnitude shorter
than that for the main sequence ones. The former are the ``silent
partners" referred to in Brown et al.\cite{Brown99C}.

Nearly all observed black hole transient X-ray sources are within
10\,kpc of the sun. Extrapolating to the entire Galaxy, a total of
$\sim 10^4$ black-hole transients with main-sequence K companions
has been suggested\cite{Brown99C}.

The lifetime of a K-star in a black hole transient X-ray source may
be the $\sim 10^{10}$\,yr lifetime of the K-star\cite{Paradijs96} 
but it is in Roche contact only $\sim 10^9$ years\cite{Brown01C}
In this
case the birth rate of the {\it observed} transient sources would be
   \be
   \lambda_K=10^4/10^9=10^{-5} {\rm per \ galaxy\ yr^{-1}}.
   \ee

We see no reason why low-mass companions should be preferred, so
we assume that the formation rate of binaries should be
independent of the ratio
   \be
   q=M_{B,i}/M_{A,i}.
   \ee
In other discussions of binaries, e.g., in Portegies Zwart \&
Yungelson\cite{Simon98}, it has often been assumed that the distribution
is uniform in $q$. This is plausible but there is no proof. Since
all primary masses $M_A$ are in a narrow interval, 20 to
$30\msun$, this means that $M_B$ is uniformly distributed between
zero and some average $M_A$, let us say $25\msun$. Then the total
rate of creation of binaries of our type is
  \be
  \lambda=\frac{25}{1.25}\lambda_K=2\times 10^{-4} {\,\rm galaxy^{-1}\ yr^{-1}}
  \label{eq7.17}
  \ee
where we have used $1.25\msun$ for the initial mass of the companion
which is stripped down to a K-star.
This is the rate of mergers of low mass black holes with neutron
stars which Bethe \& Brown\cite{Bethe98} have estimated to be
  \be
  \lambda_m \simeq 2\times 10^{-4}  {\,\rm galaxy^{-1}\ yr^{-1}}.
  \label{eq7.18}
  \ee
These mergers have been associated speculatively with short GRBs, while
formation of our binaries is supposed to lead to ``long" GRBs\cite{Fryer99}. 
We conclude that the two types of GRB should be equally frequent,
which is not inconsistent with observations. In absolute number both of our
estimates eqs.~(\ref{eq7.17}) and (\ref{eq7.18}) are
substantially larger, possibly by a factor $\sim 100$,
than the observed rate of
$10^{-7}$ galaxy$^{-1}$ yr$^{-1}$;
this is natural,
since substantial beaming is expected in GRBs produced by the
Blandford-Znajek mechanism\cite{Wijers98}. 
Although we feel our mechanism to be fairly general, it may be that the
magnetic fields required to deliver the BZ energy within a suitable time
occur in only a fraction of the He cores.
Also, from our above estimates, only the binaries with preexplosion
periods $P\lsim 1$ day would be expected to deliver enough rotational
energy to power a strong GRB. Thus the realistic $\lambda$ could 
easily come down to $\sim 10^{-5}$ galaxy$^{-1}$ yr$^{-1}$.

With our assumed distribution uniform in $q$, the silent partners
with $M_d\gsim 3\msun$ fill up nearly the entire interval; i.e.,
$\Delta q\sim 1$. For the binaries with main sequence companions
$M_d\lsim 2.5\msun$, so our interval up to $\sim 25\msun$ would
give an order of magnitude more, roughly $10^5$. However,
they will be observed only for the thermal time
Eq.~(\ref{thermal}) which is only a fraction
  \be
  F=\frac{3\times 10^7/ (M/\msun)^2}{10^{10}/(M/\msun)^{2.5}}
  =3\times 10^{-3} (M/\msun)^{1/2},
  \ee
of their total lifetime, which we take to be 
  \be
  F\sim 10^{-2}.
  \ee
This brings the birth rate of ULXs down from that of GRBs,
Eq.~(\ref{eq7.17}), to
  \be
  \lambda_{\rm ULX}\sim 2\times 10^{-6} {\rm galaxy}^{-1}\; {\rm yr}^{-1}
  \ee
in rough agreement with King et al.\cite{King01}.

In our Galaxy we see only the incipient ULX V4641 Sgr which is beginning
to cross the Herzsprung gap and GRS 1915$+$105 which is near the end
of its crossing. The latter is not beamed towards us. Both are within
12 kpc, about 1/4 of the way to the edge of the Galaxy which we take
as two dimensional. Thus, we would estimate $\sim 16 - 32$ times 
more in the entire Galaxy. With a beaming factor of 10, this would
imply $\sim 1-2$ ULXs from the SXTs in our Galaxy.

We should not forget about the two black hole binaries LMC X-3 and LMC X-1
in the Large Magellanic Cloud, which, like Cyg X-1, shine continuously,
with luminosities of $\sim 2-3\times 10^{38}$ ergs sec$^{-1}$\cite{Makishima00}
These might well be ULXs were they beamed in our direction.
Because of its low metallicity, winds in the LMC may be substantially
less than in the Galaxy, for which we argue that the strong winds during
He core burning reduce the final Fe core size in giants below the
mass necessary for later formation of high mass black holes.
Including the LMC black holes doubles our Galactic estimate and suggests
that there might be a correlation in the number of ULXs with metallicity.

\section{DISCUSSION AND CONCLUSION}
\label{conclu}

Our work here has been based on the Blandford-Znajek mechanism of
extracting rotational energies of black holes spun up by accreting
matter from a helium star. We present it using the simple circuitry of {\it
``The Membrane Paradigm"}\cite{Thorne86}. Energy
delivered into the loading region up the rotational axis of the black
hole is used to power a GRB. The energy delivered into the accretion disk
powers a SN Ib explosion.

We also discussed black-hole transient sources, high-mass black holes
with low-mass companions, as possible relics for both GRBs
and Type Ib supernova explosions, since there are indications that
they underwent mass loss in a supernova explosion. In Nova Sco
1994 there is evidence from the atmosphere of the companion star
that a very powerful supernova explosion (`hypernova') occurred.

We have shown that there is an observed correlation between orbital period
and black-hole mass in SXTs. We have modeled this
correlation as resulting from the spin of the helium star progenitor of
the black hole: if the pre-explosion orbit has a short period, the helium star
spins rapidly. This means that some part of its outer envelope is centrifugally
prevented from falling into the black hole that forms at the core. This
material is then expelled swiftly, leading to a black hole mass less
than the helium star mass.  As the orbital
period is lengthened, the centrifugal support wanes, leading to a more
massive black hole.  The reason for swift expulsion of material held up
by a centrifugal barrier is the fact that black holes formed in our
scenario naturally have high Kerr parameters (Fig.~\ref{FIG12}). This implies
that they input very high energy fluxes into their surrounding medium via
the Blandford-Znajek mechanism, and thus power both a GRB and
the expulsion of the material that does not immediately fall in.

However, because the correlation is induced between
the orbital period before explosion and the black-hole mass, its 
manifestation in the observed correlation between BH mass and present
orbital period is weakened due to
post-explosion evolution of the binaries. We therefore considered the
evolution in some detail, and for a subset of the systems were able to 
reconstruct the pre-explosion orbital periods.
The correlation between pre-explosion period and black hole mass 
(Fig.~\ref{FIG11}) is in much better agreement with our model
than the original one between present
period and black hole mass (Fig.~\ref{FIG1}). We developed a quantitative
model for the relation between period and mass, and showed that it fits the
subset of reconstructible SXT orbits.

Nova Scorpii stands out as the most extreme case of mass loss, nearly
half of the total system mass, and, therefore, a great widening
in the orbit which gets its period well beyond the gap between shrinking
and expanding orbits.
{}From Fig.~\ref{FIG11} we see that its black hole mass is far below the
polytropic line for its $M_{\rm He}=11\msun$ progenitor. We believe
that in the case of this binary a short central engine time of
several seconds was able to furnish angular momentum and energy
to the disk quickly enough to stop the infall of some of the
interior matter not initially supported by centrifugal force;
i.e., the angular momentum was provided in less than a dynamical time.
In other words, the Blandford-Znajek mechanism that drives
the GRB not only expelled
the matter initially supported for a viscous time by angular momentum,
but actually stopped the infall within a dynamical time.

Since we can also compute the Kerr parameters of the black holes formed via
our model, we find that the short-period systems should have formed black 
holes with Kerr parameters in the range 0.7--0.9. This makes them prime
candidates for energetic hypernovae and GRBs, and thus provides further
support for our earlier study in which we posited that SXTs with black-hole
primaries are the descendants of GRBs. We can now also refine this
statement: SXTs {\it with short orbital periods\/} before the formation of the
black hole have given rise to a GRB in the past.

%

We estimate the progenitors of transient sources to be formed at a
rate of 200 GEM (Galactic Events per Megayear).  Since this is
much greater than the observed rate of GRBs, there must be strong
beaming and possible selection of high magnetic fields in order to
explain the discrepancy.

We believe that there are strong reasons that a GRB must be associated
with a black hole, at least those of duration several seconds or more
discussed here. Firstly, neutrinos can deliver energy from a stellar
collapse for at most a few seconds, and sufficient power for at most
a second or two. Our quantitative estimates show that the rotating black
hole can easily supply the energy as it is braked, provided the ambient
magnetic field is sufficiently strong. The black hole also solves
the baryon pollution problem: we need the ejecta that give rise to the
GRB to be accelerated to a Lorentz factor of 100 or more, whereas the
natural scale for any particle near a black hole is less than its mass.
Consequently, we have a distillation problem of taking all the energy
released and putting it into a small fraction of the total mass.
The use of a Poynting flux from a black hole in a magnetic field\cite{bz77}
does not require the presence of much mass,
and uses the rotation energy of the black hole, so it provides
naturally clean power.

Of course, nature is extremely inventive, and we do not claim that 
all GRBs will fit into the framework outlined here. We would not
expect to see all of the highly beamed jets following from the BZ mechanism
head on, the jets may encounter some remaining hydrogen envelope in some
cases, jets from lower magnetic fields than we have considered here may
be much weaker and delivered over longer times, etc., so we speculate that
a continuum of phenomena may exist between normal supernovae and 
extreme hypernovae/GRBs.

\section*{Acknowledgments}

We would like to thank Ralph A.M.J. Wijers and Hans Bethe for the
helpful discussions.
This work is partially supported by the U.S. Department of Energy under grant
DE-FG02-88ER40388. 
CHL is supported also in part by the BK21 project of the Korea Ministry of
Education.

\appendix

\section{The Case for Case C Mass Transfer}

The proportionality of the separation $a_f$ following common envelope
evolution to $a_i$, scaled with donor mass $M_d$ makes it possible
for us to understand why the 6 binaries with shortest periods in
Table~\ref{tab1} all have K or M companions in main sequence.
(Not much is known about V406 Vulpeculae and we assume MM Velorum
to be in main sequence.)
This would be a striking coincidence, the progenitor binaries
all having black hole progenitor masses of ZAMS $20-30\msun$
and the companion ZAMS  $<2.5\msun$, unless there were some
underlying reason for this regularity. We have emphasized in the
text that in Case C mass transfer the orbital separations
for the above progenitor binaries in Roche Lobe contact
are at $\sim 1700\rsun$ ($\pm \sim 10\%$), $\sim 8$ AU, 
for ZAMS $20\msun$ black hole progenitor and $1\msun$ companion star.

The energy to remove the envelopes of the supergiants is
furnished in common envelope evolution by the drop in gravitational
binding energy of the companion, as it drops from $a_i$ to $a_f$.
This energy is thus linearly proportional to the companion mass 
which we label $M_d$. The reason that the companion masses are so
low is that the gravitational binding of the supergiant envelope
goes inversely with its radius, as $R^{-1}$, so that for $R\sim 5$ AU
(about 2/3 of the orbital separation)
the envelope binding is very low. We believe that to be the principle,
although complications enter through the variation in ZAMS masses of
the supergiants.

We were led to Case C mass transfer because in Case B mass transfer
(Roche Lobe over flow) so much of the naked He core blows away that 
the resulting
Fe core is not massive enough to form high-mass compact objects\cite{Brown01A}.

The question is, if Case C mass transfer keeps the He star clothed
during He burning so that the resulting Fe core is massive enough to collapse
into a high-mass black hole, then why wouldn't late Case B mass transfer
work? In fact, we find that very late Case B with transfer in the last
$\sim 6\times 10^4$ yrs of the He core burning, probably would work.

We should confess that our Case C mass transfer (or very late Case B)
does not have support in the results of present stellar evolution
except for the ZAMS $20\msun$ star\cite{Schaller92}.
We have based it on the regularities in the SXTs. We now turn the
argument around and say how present stellar evolution must be modified
in order to lead to these regularities.

\begin{figure}
\epsfig{file=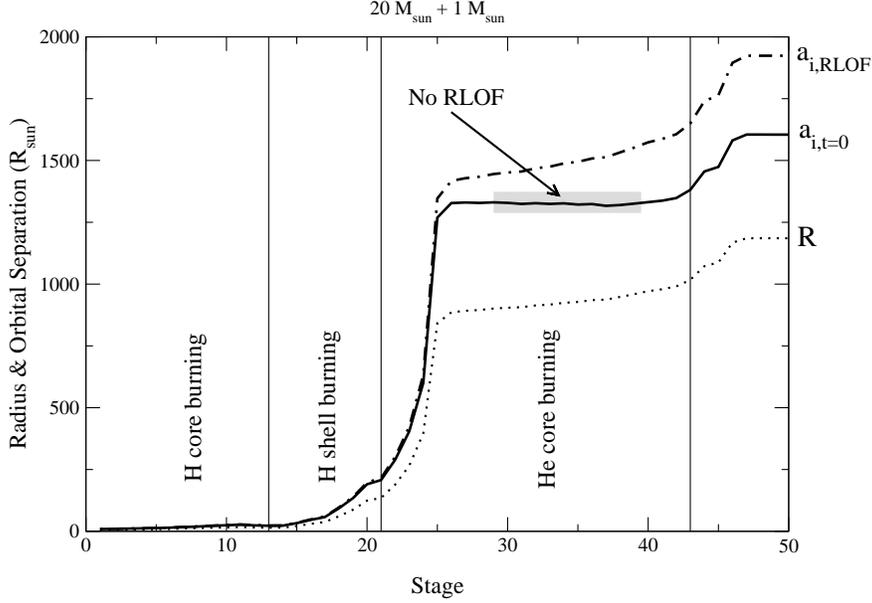,height=3.7in}
\caption{
Radius of black hole progenitors ($R$) and the
initial orbital separations ($a_i$) of the progenitors of X-ray transient
binaries with a $1\msun$ companion. 
{\bf A)} The lower dotted curves ($R$) corresponds to the radius of
the black hole progenitors taken from Schaller et al.\protect\cite{Schaller92}.
That for the
$25\msun$ star is similar but for the $30\msun$ the radius does not
increase following the end of He core burning.
{\bf B)} 
From the mass of the primary at the tabulated point
one can calculate the semimajor axis 
of a binary with a $1\msun$
secondary in which the primary fills its Roche Lobe, and this 
semimajor axis  is shown in the upper dot-dashed curve ($a_{i,{\rm RLOF}}$).
{\bf C)} The solid curves ($a_{i,t=0}$) correspond to the required initial
separations after corrections of the orbit widening due to the
wind mass loss, $a_{i,t=0}= a_{i,{\rm RLOF}}\times (M_p+M_d)/(M_{p,0}+M_d)$
where $M_p$ is the mass of the black hole progenitor at a given stage
and $M_{p,0}=20\msun$ is the ZAMS mass of the black hole progenitor. 
Primaries at the evolutionary stages marked by the shaded area
cannot fill their Roche Lobe for the first time at that stage,
but have reached their Roche Lobe at an earlier point in their
evolution.
}

\label{bltf1}
\end{figure}

We show in Fig.~\ref{bltf1} the results of the Schaller 
et al.\cite{Schaller92} stellar evolution for a ZAMS $20\msun$ star.
It is seen that the main increase in radius comes after the start
of He core burning (which begins while H shell burning is still going on).
With further He core burning there is a flattening off of the radius
versius burning stage and then a further increase in radius towards
the end of and following He core burning. Our model requires that 
mass transfer take place during this last period of increase in radius,
so that the orbital separation 
($\sim 3/2$ of the giant radius) is well
localized $a_{i,{\rm RLOF}} \sim 1700\rsun$ at the time of Roche Lobe
contact, or $a_{i,t=0}\sim 1500\rsun$ initially.

Portegies Zwart et al.\cite{Simon97} have pointed out that wind
loss from the giant preceding common envelope evolution is
important and we follow their development in identifying
the ``No RLOF" part of the curve in Fig.~\ref{bltf1}. 
Because of the wind loss the binary widens. The Roche
Lobe overflow will take place during the very rapid increase in 
radius of the giant in the beginning of He core burning, or in very
late Case B or in Case C mass transfer. The binary has widened
too much by the time the giant has reached the flat part of the
$R$ vs stage curve. In fact, it cannot transfer mass during this
stage, because it will have already come to Roche contact during
early Case B. This is made clear by the shaded area in the
solid line in Fig.~\ref{bltf1}. Brown et al.\cite{Brown01C} have
shown that the SXTs with main sequence companions can be
evolved with a $1-1.25\msun$ companion mass, so the above 
results may be directly applicable.
For higher mass companions, this shaded area becomes smaller
because the effect of the winds is smaller. This
makes the intermediate Case B mass transfer possible.
However, in this case, high-mass black holes may not form
because the Fe core is not massive enough to form high-mass compact 
objects as we discussed above\cite{Brown01A}.

Now, in fact, the curve of radius vs burning stage for the next 
massive star, of ZAMS $25\msun$, by Schaller et al.\cite{Schaller92}
does not permit Roche Lobe contact during Case C at all, the winds
having widened the binary too much by the time the giant radius begins
its last increase in late He core burning.
In the ZAMS $30\msun$ star of Schaller et al., there is no increase
in $R$ at this stage, so Case C mass transfer is not possible.

The lack of increase in $R$ for the more massive stars is due to the
cooling effect by strong wind losses. As shown by Lee et al.\cite{Lee2002}
giant progenitors as massive as $30\msun$ are necessary as progenitors 
of some of the black holes in the SXTs, especially for the binaries
with evolved companions, in order to furnish the high mass black hole
masses. These authors reduce wind losses by hand, forcing the resulting
curve of $R$ vs burning stage to look like that for a ZAMS $20\msun$ shown
in Fig.~\ref{bltf1} during the He core burning where the effect
of wind loss is important. In other words, in order to get the observed 
regularities in the evolution of SXTs, especially Eq.~(\ref{eqaf}) which 
gives the linear dependence on $M_d$ of the preexplosion separation 
of the binary, we must manufacture $R$ vs burning stage curves for which mass
transfer can be possible both early in Case B and 
in Case C. With early
Case B mass transfer, or intermediate Case B mass transfer if it occurs,
the winds during He core burning are so strong
that not enough of an Fe core is left to result in a high-mass black hole,
rather, a low-mass compact object results\cite{Brown01A}.

This story is somewhat complicated, but there have been many years of
failures in trying to evolve black holes in binaries without taking
into account the effects of binarity (mass transfer in our model)
on the evolution. On the other hand, de Kool et al.\cite{DeKool87}
had no difficulty in evolving A0620 in Case C mass transfer. The 
necessity in a similar evolution for the other black hole 
binaries was, however, not realized at that time.


\def\ApJ{{\it Astrophysical Journal}$\;$}
\def\ApJL{{\it Astrophysical Journal Letters}$\;$}
\def\AJ{{\it Astronomical Journal}$\;$}
\def\newast{{\it New Astronomy}$\;$}
\def\ARAnA{{\it Annual Reviews of Astronomy and Astrophysics}$\;$}
\def\AnA{{\it Astron. and Astrophys.}$\;$}
\def\AnAS{{\it Astron. and Astrophys. Suppl.}$\;$}
\def\MNRAS{{\it Mon. Not. of Royal. Astron. Soc.}$\;$}
\def\PASP{{\it Pub. of the Astron. Soc. of the Pac.}$\;$}

\end{document}